\documentclass[aps,prb,reprint]{revtex4-1}

\usepackage{graphicx}
\usepackage{enumerate}
\usepackage{hyperref}
\usepackage{textcomp}
\usepackage{amsmath}
\usepackage{amssymb}
\usepackage{pgf}
\usepackage{bm}
\usepackage{epsfig}
\usepackage{figsize}

\begin{document}

\title{Competing electronic orders on a heavily doped honeycomb lattice\\ with enhanced exchange coupling}

\author{Song-Jin O}
\email{sj.o@ryongnamsan.edu.kp}
\author{Yong-Hwan Kim, Ok-Gyong Pak, Kum-Hyok Jong, Chol-Won Ri}
\author{Hak-Chol Pak}
\affiliation{Faculty of Physics, Kim Il Sung University, Ryongnam-Dong, Taesong District, Pyongyang, Democratic People's Republic of Korea}
\date{\today}

\begin{abstract}
Motivated by recent discovery of correlated insulating and superconducting behavior in twisted bilayer graphene, we revisit graphene's honeycomb lattice doped close to the van Hove singularity, using the truncated unity functional renormalization group approach. We consider an extended Hubbard model on the honeycomb lattice including on-site and nearest-neighbor Coulomb repulsions, and nearest-neighbor ferromagnetic exchange and pair hopping interactions. By varying the strength of the nearest-neighbor exchange coupling and Coulomb repulsion as free parameters, we present rich ground-state phase diagrams which contain the spin-triplet $f$-wave and spin-singlet chiral $d$-wave superconducting phases, the commensurate and incommensurate spin- and charge-density-wave phases, and the ferromagnetic phase. In the absence of the exchange coupling and for small value of the nearest-neighbor repulsion, the four-sublattice spin-density-wave phase is generated right around the van Hove filling, while the chiral $d$-wave superconductivity emerges slightly away from it. Surprisingly, the chiral $d$-wave superconductivity is strongly suppressed by weak nearest-neighbor exchange coupling in our calculations. We argue that this suppression might be one of the reasons why the chiral superconductivity proposed for doped graphene has not yet been observed experimentally.
\end{abstract}


\maketitle

\section{Introduction}

The recent discoveries of correlated insulating and superconducting behaviors in twisted bilayer graphene (TBG) \cite{ref01, ref02, ref03} have generated great interest in the study of graphene-based systems. Since this two-dimensional heterostructure has an unprecedented tunability, the TBG can serve as a new platform to study the correlated electron systems like high-temperature superconductors. A considerable theoretical effort has been devoted to the study on the pairing mechanism and symmetry for the superconducting state of TBG. Many of them have pointed to a chiral $d$-wave superconductivity (SC) supporting the nontrivial topology as the leading instability in the pairing channel \cite{ref04, ref05, ref06, ref07, ref08, ref09, ref10, ref57, ref58, ref59, ref60}. The discovery of SC in TBG has motivated us to revisit the issue of the possibility of unconventional SC in a doped single layer graphene.

Chiral SC is characterized by the phase of the superconducting order parameter winding by multiples of $2\pi$ around the Fermi surface (FS), breaking parity and time-reversal symmetry \cite{ref11, ref12}. Generally, it is formed by a complex linear combination of two order parameters belonging to a two-dimensional irreducible representation of the point group of the crystal. For example, the chiral $d$-wave superconducting state (SC state) predicted for the honeycomb lattice originates from two nodal $d$-wave states that are degenerate by $C_{6v}$ symmetry of the lattice. Those two degenerate states can construct the chiral $d$-wave SC via a complex linear combination, thus giving a full gap and the energy gain (for a review, see Ref. \onlinecite{ref12}).

The chiral $d$-wave SC driven by the electron-electron interaction has been theoretically proposed for graphene's honeycomb lattice near half filling \cite{ref13, ref14, ref15, ref16, ref17}. However, unrealistically high values of predicted transition temperatures \cite{ref13, ref15, ref17}, and large value of the antiferromagnetic exchange coupling needed for the emergence of the SC state \cite{ref14}, imply that these results are either unreliable or not appropriate for graphene. It has been argued that the SC state can be destroyed by quantum fluctuations in charge or spin channel for the Hubbard model on the honeycomb lattice \cite{ref14, ref16}. Small electronic density of states near half-filling and weak phonon effect are disadvantageous for developing the superconducting order (SC order) in weakly doped graphene.

In the case of graphene, the electron-driven SC state is most likely realized when doped to the vicinity of the van Hove singularity (VHS). Near the VHS, a combination of the logarithmically divergent density of states and the approximate nesting of the FS strongly enhances the effect of interactions \cite{ref18, ref19, ref20}, which can lead to the emergence of a variety of ordered states \cite{ref21} at relatively high temperatures. In previous works on graphene near the VHS filling, various electronic instabilities were analyzed using the mean-field theory \cite{ref29, ref31, ref34}, random phase approximation \cite{ref22, ref33}, quantum Monte Carlo (QMC) \cite{ref25, ref32}, variational \cite{ref24, ref27}, and renormalization group (RG) \cite{ref11, ref23, ref26} approaches.

Both the random phase approximation \cite{ref22} and the perturbative RG studies \cite{ref11} have predicted that the electron-driven chiral $d$-wave SC could emerge upon doping graphene towards or onto the VHS. Several calculation results using the $N$-patch functional renormalization group (FRG) \cite{ref23}, Grassmann tensor product state \cite{ref24}, and finite-temperature determinantal QMC \cite{ref25} approaches also support the chiral $d$-wave SC in the vicinity of the VHS, while the singular-mode FRG calculation \cite{ref26} reports it for doping away from the VHS. Another instability towards the spin-triplet $p$-wave SC near the VHS filling has been found in a study using variational cluster approximation and cellular dynamical mean-field theory \cite{ref27}, and a dynamic cluster approximation calculation \cite{ref28}.

The charge and magnetic instabilities have also been analyzed in previous works on graphene doped close to the VHS. A Pomeranchuk instability has been reported in the calculations by mean-field theory \cite{ref29} and generalized Pomeranchuk method \cite{ref30}. There have been several works that found a topologically nontrivial chiral spin-density-wave (SDW) order at the VHS filling by using the mean-field theory \cite{ref31}, the singular-mode FRG \cite{ref26}, the finite-temperature determinantal QMC \cite{ref25}, and a combination of exact diagonalization, density matrix RG, and variational Monte Carlo methods \cite{ref32}. The random phase approximation \cite{ref33} and the mean-field theory \cite{ref34} have reported another SDW order for doping at the VHS. Thus, there exist remarkable diversity and discrepancy regarding predicted electronic instabilities for the honeycomb lattice near the VHS filling.

On the other hand, in the studies above, the electrons on the graphene's honeycomb lattice are mostly described by both the Hubbard model with on-site and nearest-neighbor Coulomb interactions and the $t$-$J$ model which is derived from the Hubbard model with strong on-site repulsion. A FRG study \cite{ref14} employed the Hubbard model with additional antiferromagnetic exchange interaction, but to our knowledge, there have been no previous studies on the doped honeycomb lattice where the ferromagnetic exchange interaction was taken into account. It is well known that the expansion of the Coulomb interaction Hamiltonian in the localized Wannier orbitals produces the ferromagnetic exchange couplings and pair-hopping terms between neighboring sites (the detailed description of it can be found in Appendix \ref{appendA}). We argue that the combination of ferromagnetic exchange and pair hopping should be involved into the bare interaction, though it is very weak and may be canceled by the antiferromagnetic exchange couple generated by virtual hopping processes for strong on-site repulsion. A rich ground-state phase diagram for a half-filled honeycomb lattice has been created by the extended Hubbard model involving the ferromagnetic exchange interaction \cite{ref35}.

The above-mentioned facts show that a consistent picture of possible electronic instabilities on the honeycomb lattice described by the Hubbard model is still lacking, probably due to competition between several ordering tendencies upon varying the interaction parameters or the doping. Hence, it appears promising and necessary to investigate the system using a reliable and unbiased method that allows us to alter these parameters over a wide range.

Here, we employ the recently developed truncated unity functional renormalization group (TUFRG) approach \cite{ref36} with a high momentum resolution to study the competing electronic orders on the honeycomb lattice near the VHS filling with a focus on the effect of the nearest-neighbor exchange interaction. Taking into account the exponential decay of the ferromagnetic exchange coupling with the interatomic distance and the strong screening of the density-density interaction due to large value of the density of states near the VHS, we consider the Hubbard model including the on-site repulsion $U$, the nearest-neighbor repulsion $V$, and the nearest-neighbor ferromagnetic exchange coupling $J$. Based on it, we build the tentative phase diagrams in the space of the nearest-neighbor repulsion $V$ and the doping level $\delta$ for the fixed value of $U$ and several typical values of $J$, which would provide a comprehensive picture of possible ordered ground states and a reasonable description for the effects of the interaction parameters on the ordering tendencies.

Our main result are summarized as follows. In the absence of the exchange coupling $J$ and for small nearest-neighbor repulsion $V$, the four-sublattice SDW phase is generated right around the VHS, while the chiral $d$-wave SC emerges slightly away from it, which is similar to the result in Ref. \onlinecite{ref26}. Upon increasing $V$, the spin-triplet $f$-wave SC becomes dominant below the VHS. If $V$ is further increased, the charge-density-wave (CDW) state with broken $\pi /3$-rotation symmetry and a charge transfer from sublattice $A$ to $B$ (or vice versa) will be favored for all doping levels studied. The incommensurate SDW and CDW orders are also found in some regions of the phase diagram. Remarkably, upon involving weak exchange coupling of $J \approx 0.14{\rm{eV}}$, the chiral $d$-wave SC completely disappears from our phase diagram. For an artificially large value of $J$, the phase diagrams have simple structures consisting of both the CDW and the ferromagnetic phases.

This paper is organized as follows. In Sec.~\ref{sec2} we specify the model Hamiltonian and give a description of the TUFRG scheme. In Sec.~\ref{sec3} we present and analyze the schematic phase diagrams for electrons subjected to the on-site repulsion, the nearest-neighbor repulsion, and exchange interaction. Finally, in Sec.~\ref{sec4} we draw our conclusions.

\section{Model and Method} \label{sec2}

\subsection{Extended Hubbard model}\label{sec2A}
We study spin-1/2 electrons on the honeycomb lattice doped close to the VHS which are described by an extended Hubbard model including ferromagnetic exchange interaction. The noninteracting part of the model is represented by a tight-binding Hamiltonian with nearest-neighbor and next-nearest-neighbor hoppings,
\begin{eqnarray} \label{eq01}
\begin{split}
H_0  = & - t\sum\limits_{\left\langle {iA,jB} \right\rangle ,\sigma } {(c_{iA\sigma }^\dag c_{jB\sigma } }  + {\rm{H}}{\rm{.c}}{\rm{.}}) \\
& - t'\sum\limits_{\left\langle {\left\langle {io,jo} \right\rangle } \right\rangle ,o,\sigma } {(c_{io\sigma }^\dag c_{jo\sigma } }  + {\rm{H}}{\rm{.c}}{\rm{.}}) - \mu N_e,
\end{split}
\end{eqnarray}
where the operator $c_{io\sigma }^\dag$ ($c_{io\sigma }$) creates (annihilates) an electron at lattice site $i$ with spin polarity $\sigma$ in sublattice $o$, $\left\langle {iA,jB} \right\rangle$ ($\left\langle {\left\langle {io,jo} \right\rangle } \right\rangle$) denotes nearest-neighbor (next-nearest-neighbor) bonds, $\mu$ is the chemical potential, and $N_e$ is the total electron number operator. The doping level is controlled by the chemical potential and defined by $\delta  = n_e  - 1$ where $n_e$ is the number of electrons per site. These parameters have the values of $\mu _{{\rm{VHS}}}  = t + 2t',\delta _{{\rm{VHS}}}  = 0.25$  at the VHS filling.
The interaction part of the model is given by
\begin{eqnarray} \label{eq02}
\begin{split}
H_{{\rm{int}}}  = {\kern 3pt} & U \sum\limits_{i,o} {n_{io \uparrow } } n_{io \downarrow }  + V\sum\limits_{\left\langle {iA,jB} \right\rangle } {\sum\limits_{\sigma ,\sigma '} {n_{iA\sigma } } } n_{jB\sigma '}\\ 
& + J\sum\limits_{\left\langle {iA,jB} \right\rangle } {\sum\limits_{\sigma ,\sigma '} {c_{iA\sigma }^ \dag  c_{jB\sigma '}^ \dag  } } c_{iA\sigma '} c_{jB\sigma }  \\
& + J\sum\limits_{\left\langle {iA,jB} \right\rangle } {(c_{iA \uparrow }^ \dag  c_{iA \downarrow }^ \dag  } c_{jB \downarrow } c_{jB \uparrow }  + {\rm{H}}{\rm{.c}}{\rm{.}}),
\end{split}
\end{eqnarray}
where $n_{io\sigma }  = c_{io\sigma }^ \dag  c_{io\sigma } $ is the local electron density operator for spin polarity  $\sigma$, and the terms in Eq. (\ref{eq02}) represent the on-site and nearest-neighbor density-density interactions, the nearest-neighbor ferromagnetic exchange interaction, and the nearest-neighbor pair hopping. The derivation of this interaction Hamiltonian is presented in Appendix \ref{appendA}.

We allow the extended ranges of the parameters $V$ and $J$ to investigate their effects on the ground state of the heavily doped honeycomb lattice, but not constrained by the actual values of graphene. We take $t = 2.8{\rm{eV}}, t' = 0.1{\rm{eV}}, U = 3.6t$, as suggested in Ref. \onlinecite{ref37} and used in Ref. \onlinecite{ref26}. In our calculations we have considered the parameters $\delta$ and $V$ in the ranges of $[0.19,0.31]$ and $[0,3t]$, respectively.

\subsection{Truncated unity functional\\ renormalization group method}\label{sec2B}
Since graphene has the bandwidth ($\sim 17{\rm{eV}}$) of the order of the interaction scale ($\sim 10{\rm{eV}}$) and does not exhibit any Mott insulating behavior at half filling, it can be thought to be in the intermediate coupling regime. The FRG method \cite{ref38, ref39, ref40} is known to operate most reliably for intermediate coupling. It serves as an unbiased tool for investigating correlated electron systems and accounts for the competition and mutual interplay between different channels because it takes into account an infinite sum of all possible one-loop diagrams including the vertex corrections between the particle-particle and particle-hole channels, on equal footing \cite{ref40, ref54}. In fact, it has already been used to investigate the FS instability in doped graphene \cite{ref23, ref26}.

As a modified version of the FRG, the TUFRG approach \cite{ref36} is based on the exchange parametrization FRG \cite{ref41} and the singular-mode FRG \cite{ref26} approaches. It has the advantages that it allows for a high speed calculation with high momentum resolution and an efficient parallelization on a large number of computer nodes \cite{ref42}. The TUFRG method has been applied to the analysis of the electronic instabilities for the half-filled honeycomb lattice \cite{ref35, ref43} and strained graphene \cite{ref44}. We consider the system with spin-SU(2) symmetry and calculate the effective interactions in the orbital picture of TUFRG. As the regulator for infrared divergences, the $\Omega$ scheme \cite{ref41} is employed, in which the bare propagator $G_{oo'}^0 (\omega ,{\bf{k}})$ for orbital indices $o,o'$ is modified by energy scale $\Omega$ as
\begin{equation}
\nonumber
G_{oo'}^0 (\omega ,{\bf{k}}) \to G_{oo'}^{0,\Omega } (\omega ,{\bf{k}}) = \frac{{\omega ^2 }}{{\omega ^2  + \Omega ^2 }}G_{oo'}^0 (\omega ,{\bf{k}}).
\end{equation}
The regulated propagator $G^{0,\Omega }$ leads the generating functional of one-particle-irreducible vertex functions to be scale dependent as well, $\Gamma  \to \Gamma ^\Omega$. By differentiating $\Gamma ^\Omega$ with respect to $\Omega$, one can obtain the functional flow equation which is then Taylor expanded to produce an infinite hierarchy of flow equations for the vertex functions.

In numerical implementation, the hierarchy has to be truncated at a certain order. In the study of ground-state properties, one generally use a truncation in which all $n$-point vertices with $n \ge 6$ are set to zero, and the self-energy correction and the frequency dependence of the 4-point vertex function are neglected. Such an approximation has proven to provide reliable results for many two-dimensional systems \cite{ref39, ref40}, and the truncation can be justified by the reasoning of Salmhofer and Honerkamp \cite{ref45}. Assuming weak to moderate 4-point vertex and the absence of higher-order, i.e., $n$-point ($n \ge 6$) vertices at a bare level, it was shown that for high-energy scales of renormalization, where the 4-point vertex $\gamma^{(4)}$ (or the effective interaction $V$ in this paper) is still relatively small, the contributions of higher-order vertices are likewise small as they are developed by higher-order terms of $\gamma^{(4)}$. At intermediate scales, the contributions of higher-order vertices remain small for the case of sufficiently smooth and curved FS, even though the scale-dependent $\gamma^{(4)}$ is no longer small, according to a phase-space argument. Only at low-energy scales, where $\gamma^{(4)}$ gets diverged, the smallness of relevant phase space cannot suppress the higher-order contributions, and the RG flow has to be stopped \cite{ref40}. So the FRG with this truncation is expected to be appropriate to identify the many-particle instabilities for systems in the intermediate-coupling regime like graphene. For further discussion, we refer to the articles [\onlinecite{ref39}, \onlinecite{ref45}].

Within this approximation the 4-point part of the generating functional for spin-SU(2)-invariant systems can be represented by the effective interactions $V^\Omega$ and the Grassmann variables $\bar \psi ,\psi$ as
\begin{eqnarray} \label{eq03}
\begin{split}
\Gamma ^{\Omega ,(4)} [\bar \psi ,\psi ] = & \frac{1}{2}\int {d\xi _1 }  \cdots d\xi _4 {\kern 3pt}  V_{o_1 o_2 ,o_3 o_4 }^\Omega  ({\bf{k}}_1 ,{\bf{k}}_2 ;{\bf{k}}_3 ,{\bf{k}}_4 )\\
& \times \delta (k_1  + k_2  - k_3  - k_4 )\\
& \times \sum\limits_{\sigma ,\sigma '} {\bar \psi _\sigma  (\xi _1 )} \bar \psi _{\sigma '} (\xi _2 )\psi _{\sigma '} (\xi _4 )\psi _\sigma  (\xi _3).
\end{split}
\end{eqnarray}
Here $k_i  = (\omega _i ,{\bf{k}}_i )$ and $\xi _i  = (\omega _i ,{\bf{k}}_i ,o_i )$ are multi-indices comprising a Matsubara frequency $\omega _i$, wave vector ${\bf{k}}_i$, and orbital (sublattice) index $o_i$ and we have introduced the abbreviation $\int {d\xi _i }  = \int {\frac{{d{\bf{k}}_i }}{{S_{BZ} }}\frac{1}{\beta }} \sum\nolimits_{\omega _i } {\sum\nolimits_{o_i }}$ with the Brillouin zone (BZ) area $S_{BZ}$ and inverse temperature $\beta$. The flow equation of the effective interaction reads \cite{ref45, ref46}
\begin{eqnarray} \label{eq04}
\frac{d}{{d\Omega }}V^\Omega   = J^{{\rm{pp}}}(\Omega)  + J^{{\rm{ph,cr}}}(\Omega)  + J^{{\rm{ph,d}}}(\Omega),
\end{eqnarray}
where the expressions for $J^{{\rm{pp}}}(\Omega)$, $J^{{\rm{ph,cr}}}(\Omega)$ \footnote{There is an error in expression for $J^{{\rm{ph,cr}}}$ in Eq. (5) of Ref. \cite{ref35}, which was revised in Eq. (\ref{eq06}) of this paper. In addition, there is an another error in expression for $T_{ob}({\bf{k}})$ in Eq. (43) of the reference, which should be corrected as $T_{ob}({\bf{k}}) = \left({\sqrt 2}\right)^{-1} \left(\begin{array}{cc} \frac{d({\bf{k}})}{|d({\bf{k}})|} \quad \frac{d({\bf{k}})}{|d({\bf{k}})|}\\ -1 {\kern 23pt} 1{\kern 6pt} \end{array}\right)$.} and $J^{{\rm{ph,d}}}(\Omega)$ are
\begin{eqnarray} \label{eq05}
\begin{split}
	J_{o'_1 o'_2 ,  o_1 o_2 }^{{\rm{pp}}(\Omega )} & ({\bf{k}}'_1 ,{\bf{k}}'_2 ;{\bf{k}}_1 ,{\bf{k}}_2 ) =
	- \sum\limits_{\mu ,\mu '} {\sum\limits_{\nu ,\nu '} {\int {dp} } }\\
	\frac{d}{{d\Omega }} & \left[ {G_{\mu \nu }^{0,\Omega } (\omega ,{\bf{p}} + {\bf{k}}'_1  + {\bf{k}}'_2 )G_{\mu '\nu '}^{0,\Omega } ( - \omega , - {\bf{p}})} \right]\\
	& \times V_{o'_1 o'_2 ,\mu \mu '}^\Omega  ({\bf{k}}'_1 ,{\bf{k}}'_2 ;{\bf{p}} + {\bf{k}}'_1  + {\bf{k}}'_2 , - {\bf{p}})\\
	& \times V_{\nu \nu ',o_1 o_2 }^\Omega  ({\bf{p}} + {\bf{k}}'_1  + {\bf{k}}'_2 , - {\bf{p}};{\bf{k}}_1 ,{\bf{k}}_2 ),
\end{split}
\end{eqnarray}	
\begin{eqnarray} \label{eq06}
\begin{split}
	J_{o'_1 o'_2 ,o_1 o_2 }^{{\rm{ph,cr}}(\Omega )} & ({\bf{k}}'_1 ,{\bf{k}}'_2 ;{\bf{k}}_1 ,{\bf{k}}_2 ) = 
	- \sum\limits_{\mu ,\mu '} {\sum\limits_{\nu ,\nu '} {\int {dp} } } \\
	\frac{d}{{d\Omega }} & \left[ {G_{\mu \nu }^{0,\Omega } (\omega ,{\bf{p}} + {\bf{k}}'_1  - {\bf{k}}_2 )G_{\nu '\mu '}^{0,\Omega } (\omega ,{\bf{p}})} \right]\\
	& \times V_{o'_1 \mu ',\mu o_2 }^\Omega  ({\bf{k}}'_1 ,{\bf{p}};{\bf{p}} + {\bf{k}}'_1  - {\bf{k}}_2 ,{\bf{k}}_2 ) \\
	& \times V_{\nu o'_2 ,o_1 \nu '}^\Omega  ({\bf{p}} + {\bf{k}}'_1  - {\bf{k}}_2 ,{\bf{k}}'_2 ;{\bf{k}}_1 ,{\bf{p}}),\\
\end{split}
\end{eqnarray}
\begin{eqnarray} \label{eq07}
\begin{split}
	J_{o'_1 o'_2 ,o_1 o_2 }^{{\rm{ph,d}}(\Omega )} & ({\bf{k}}'_1 ,{\bf{k}}'_2 ;{\bf{k}}_1 ,{\bf{k}}_2 ) =  - \sum\limits_{\mu ,\mu '} {\sum\limits_{\nu ,\nu '} {\int {dp} } } \\
	\frac{d}{{d\Omega }} & \left[ {G_{\mu \nu }^{0,\Omega } (\omega ,{\bf{p}} + {\bf{k}}'_1  - {\bf{k}}_1 )G_{\nu '\mu '}^{0,\Omega } (\omega ,{\bf{p}})} \right] \\
	\times & \left[ V_{o'_1 \mu ',\mu o_1 }^\Omega \right.({\bf{k}}'_1 ,{\bf{p}};{\bf{p}} + {\bf{k}}'_1  - {\bf{k}}_1 ,{\bf{k}}_1 ) \\
	&  \hspace{1pc} \times V_{\nu o'_2 ,\nu 'o_2 }^\Omega  ({\bf{p}} + {\bf{k}}'_1  - {\bf{k}}_1 ,{\bf{k}}'_2 ;{\bf{p}},{\bf{k}}_2 ) \\
	& + V_{o'_1 \mu ',o_1 \mu }^\Omega  ({\bf{k}}'_1 ,{\bf{p}};{\bf{k}}_1 ,{\bf{p}} + {\bf{k}}'_1  - {\bf{k}}_1 ) \\
	& \hspace{1pc} \times V_{\nu o'_2 ,o_2 \nu '}^\Omega  ({\bf{p}} + {\bf{k}}'_1  - {\bf{k}}_1 ,{\bf{k}}'_2 ;{\bf{k}}_2 ,{\bf{p}}) \\
	& - 2V_{o'_1 \mu ',o_1 \mu }^\Omega  ({\bf{k}}'_1 ,{\bf{p}};{\bf{k}}_1 ,{\bf{p}} + {\bf{k}}'_1  - {\bf{k}}_1 ) \\
	& \hspace{1pc} \times \left. V_{\nu o'_2 ,\nu 'o_2 }^\Omega  ({\bf{p}} + {\bf{k}}'_1  - {\bf{k}}_1 ,{\bf{k}}'_2 ;{\bf{p}},{\bf{k}}_2 ) \right],
\end{split}
\end{eqnarray}
with the shorthand notation $\int {dp}  = \int {\frac{{d{\bf p}}}{{S_{BZ} }}} \frac{1}{\beta }\sum \nolimits_\omega$  and implicit constraint ${\bf{k}}'_1  + {\bf{k}}'_1  = {\bf{k}}_1  + {\bf{k}}_2$. The effective interaction is obtained by integrating Eq. (\ref{eq04}) with respect to energy scale $\Omega$:
\begin{eqnarray} \label{eq08}
\begin{split}
V^\Omega   = & V^{(0)}  + \int_{\Omega _0 }^\Omega  {d\Omega '} J^{{\rm{pp}}} (\Omega ') \\
& + \int_{\Omega _0 }^\Omega  {d\Omega '} J^{{\rm{ph,cr}}} (\Omega ') + \int_{\Omega _0 }^\Omega  {d\Omega '} J^{{\rm{ph,d}}} (\Omega ') \\
= &  V^{(0)}  + \Phi ^{{\rm{pp}}} (\Omega ) + \Phi ^{{\rm{ph,cr}}} (\Omega ) + \Phi ^{{\rm{ph,d}}} (\Omega )
\end{split}
\end{eqnarray}
Here $\Omega_0 $ is the initial value of $\Omega$, $V^{(0)}  \equiv V^{\Omega _0 } $ is the initial bare interaction, and, e.g., $\Phi ^{{\rm{pp}}} (\Omega) = \int_{\Omega_0 }^\Omega  {d\Omega'} J^{{\rm{pp}}} (\Omega')$ is the single-channel coupling function.

Three bosonic propagators are defined by projecting three single-channel coupling functions onto three channels, i.e., the particle-particle, crossed particle-hole and direct particle-hole channels (a more detailed description of these projections is contained in Appendix \ref{appendB}):
\begin{eqnarray} \label{eq09}
\begin{split}
& P^\Omega   = {\rm{\hat P}}[\Phi ^{{\rm{pp}}} (\Omega )], \\
& C^\Omega   = {\rm{\hat C}}[\Phi ^{{\rm{ph,cr}}} (\Omega )],
D^\Omega = {\rm{\hat D}}[\Phi ^{{\rm{ph,d}}}(\Omega)].
\end{split}
\end{eqnarray}
They have matrix structures and depend only on one momentum in contrast to the effective interaction depending on three momenta. For example, the bosonic propagator $P^\Omega$ is a matrix that depends on the momentum transfer $\bf{q}$ and contains the elements $P_{o_1 o_2 m,o_3 o_4 n}^\Omega  ({\bf{q}})$ with sublattice indices $o_1 \sim o_4$ and basis indices $m$, $n$. Since the projections lead to the truncation in expansion of the single-channel coupling functions, the inverse projections of Eq. (\ref{eq09}) can only give approximate results for the coupling functions:
\begin{eqnarray} \label{eq10}
\begin{split}
 & \Phi ^{{\rm{pp}}} (\Omega ) \approx {\rm{\hat P}}^{ - 1} [P^\Omega  ], \\
 & \Phi ^{{\rm{ph,cr}}} (\Omega ) \approx {\rm{\hat C}}^{ - 1} [C^\Omega  ],
 \Phi ^{{\rm{ph,d}}} (\Omega ) \approx {\rm{\hat D}}^{ - 1} [D^\Omega  ].
\end{split}
\end{eqnarray}
One can also project the effective interaction $V^\Omega$ onto the three channels as
\begin{eqnarray} \label{eq11}
\begin{split}
V^{\rm{P}} (\Omega ) = {\rm{\hat P}}[V^\Omega  ],V^{\rm{C}} (\Omega ) = {\rm{\hat C}}[V^\Omega  ],V^{\rm{D}} (\Omega ) = {\rm{\hat D}}[V^\Omega ],\hspace{1.5pc}
\end{split}
\end{eqnarray}
whose inverse projections are (a more detailed description of these projection matrices is found in Appendix \ref{appendC})
\begin{eqnarray} \label{eq12}
\begin{split}
V^\Omega   \approx {\rm{\hat P}}^{ - 1} [V^{\rm{P}} (\Omega )] \approx {\rm{\hat C}}^{ - 1} [V^{\rm{C}} (\Omega )] \approx {\rm{\hat D}}^{ - 1} [V^{\rm{D}} (\Omega )].\hspace{1.5pc}
\end{split}
\end{eqnarray}

Taking the derivative of $P^\Omega  ,C^\Omega$ and $D^\Omega$ with respect to $\Omega$ one can derive the flow equations for the bosonic propagators:
\begin{eqnarray} \label{eq13}
\begin{split}
\frac{d}{{d\Omega }}P^\Omega &  = \frac{d}{{d\Omega }}{\rm{\hat P}}[\Phi ^{{\rm{pp}}} (\Omega )] 
= {\rm{\hat P}}\left[ {\frac{d}{{d\Omega }}\Phi ^{{\rm{pp}}} (\Omega )} \right] \\
& = {\rm{\hat P[}}J^{{\rm{pp}}} (\Omega ){\rm{]}},\\
\frac{d}{{d\Omega }}C^\Omega &  = {\rm{\hat C}}[J^{{\rm{ph,cr}}} (\Omega )],\frac{d}{{d\Omega }}D^\Omega   = {\rm{\hat D}}[J^{{\rm{ph,d}}} (\Omega)].
\end{split}
\end{eqnarray}
Plugging Eqs. (\ref{eq05})$-$(\ref{eq07}) into Eq. (\ref{eq13}), and representing $V^\Omega$ in terms of projection matrices $V^{\rm{P}}(\Omega),V^{\rm{C}}(\Omega)$ and $V^{\rm{D}}(\Omega)$ according to Eq. (\ref{eq12}), we arrive at the ultimate flow equations for the bosonic propagators \cite{ref36}:
\begin{eqnarray} \label{eq14}
\begin{split}
\frac{{dP^\Omega  ({\bf{q}})}}{{d\Omega }} = & V^{{\rm{P}}(\Omega )} ({\bf{q}})\dot \chi ^{{\rm{pp}}} ({\bf{q}})V^{{\rm{P}}(\Omega )} ({\bf{q}}), \\
\frac{{dC^\Omega  ({\bf{q}})}}{{d\Omega }} = & V^{{\rm{C}}(\Omega )} ({\bf{q}})\dot \chi ^{{\rm{ph}}} ({\bf{q}})V^{{\rm{C}}(\Omega )} ({\bf{q}}), \\
\frac{{dD^\Omega  ({\bf{q}})}}{{d\Omega }} = & [V^{{\rm{C}}(\Omega )} ({\bf{q}}) - V^{{\rm{D}}(\Omega )} ({\bf{q}})]\dot \chi ^{{\rm{ph}}} ({\bf{q}})V^{{\rm{D}}(\Omega)} ({\bf{q}}) \\
& + V^{{\rm{D}}(\Omega )} ({\bf{q}})\dot \chi ^{{\rm{ph}}} ({\bf{q}})[V^{{\rm{C}}(\Omega )} ({\bf{q}}) - V^{{\rm{D}}(\Omega )} ({\bf{q}})],
\end{split}
\end{eqnarray}
with
\begin{eqnarray} \label{eq15}
\begin{split}
& \dot \chi _{o'_1 o'_2 m,o_1 o_2 n}^{{\rm{pp}}} ({\bf{q}}) = - \int {dk} f_m ({\bf{k}})f_n^* ({\bf{k}}) \\
& \hspace{1.5pc} \times \frac{d}{{d\Omega }}[G_{o'_1 o_1 }^{0,\Omega } (\omega ,{\bf{k}} + {\bf{q}})G_{o'_2 o_2 }^{0,\Omega } ( - \omega , - {\bf{k}})], \\
& \dot \chi _{o'_1 o'_2 m,o_1 o_2 n}^{{\rm{ph}}} ({\bf{q}}) = - \int {dk} f_m ({\bf{k}})f_n^* ({\bf{k}}) \\
& \hspace{1.5pc} \times \frac{d}{{d\Omega }}[G_{o'_1 o_1 }^{0,\Omega } (\omega ,{\bf{k}} + {\bf{q}})G_{o_2 o'_2 }^{0,\Omega } (\omega ,{\bf{k}})].
\end{split}
\end{eqnarray}

\subsection{Symmetries and Order Parameters}\label{sec2C}
The honeycomb lattice has $C_{6v}$ point-group symmetry. This symmetry leads to symmetry relations for the Bloch states and the effective interaction, yielding the relations between the bosonic propagators with different momentum arguments. By these relations, the bosonic propagators in the whole BZ can be obtained from those within the irreducible region of the BZ, which reduces the computational effort to 1/12. In our calculation we use the plane-wave basis $f_m ({\bf{k}}) = e^{i{\bf{R}}_m  \cdot {\bf{k}}}$ for the expansion of the single-channel coupling functions in terms of the bosonic propagators. Then we can derive the explicit symmetry relations for the bosonic propagators.

Let us consider a symmetry operation $\hat G = (Q|{\bf{t}})$, i.e., a rotation $Q$ followed by shift $\bf{t}$. Under this operation the atom of sublattice $o$ in the unit cell at the origin is moved to the site of sublattice $\tilde o$ in the unit cell at the position ${\bf{u}}_o$. It can be represented by
\begin{eqnarray} \label{eq16}
Q{\bf{d}}_o  + {\bf{t}} = {\bf{u}}_o  + {\bf{d}}_{\tilde o},
\end{eqnarray}
where ${\bf{d}}_o$ is the relative position of the sublattice $o$ and ${\bf{u}}_o$ is one of the Bravais lattice vectors. The symmetry operation $\hat G$ yields the following symmetry relations for three bosonic propagators \cite{ref35}:
\begin{eqnarray} \label{eq17}
\begin{split}
P(\text{or } C &, D)_{\tilde o_1 ,\tilde o_2 ,Q{\bf{R}}_m  + {\bf{u}}_{o_1 }  - {\bf{u}}_{o_2 } ;\tilde o_3 ,\tilde o_4 ,Q{\bf{R}}_n  + {\bf{u}}_{o_3 }  - {\bf{u}}_{o_4 } }^\Omega  (Q{\bf{q}}) \hspace{1.5pc}\\
& = e^{ - iQ{\bf{q}} \cdot ({\bf{u}}_{o_1 }  - {\bf{u}}_{o_3 } )} P(\text{or } C,D)_{o_1 o_2 m,o_3 o_4 n}^\Omega  ({\bf{q}}).
\end{split}
\end{eqnarray}

In addition, the effective interactions have another symmetries, i.e., the particle-hole symmetry and the remnant of antisymmetry of Grassmann variables \cite{ref41}, which lead to the following relations \cite{ref35}:
\begin{eqnarray} \label{eq18}
\begin{split}
& P(\text{or } C,D)_{o_1 o_2 m,o_3 o_4 n}^\Omega  ({\bf{q}}) \\
& \hspace{2pc} = [P(\text{or } C,D)_{o_3 o_4 n,o_1 o_2 m}^\Omega  ({\bf{q}})]^* \hspace{3pc}
\end{split}
\end{eqnarray}
for the particle-hole symmetry and
\begin{eqnarray} \label{eq19}
\begin{split}
P_{o_1 o_2 m,o_3 o_4 n}^\Omega  ({\bf{q}}) = & e^{i{\bf{q}} \cdot ({\bf{R}}_n  - {\bf{R}}_m )} P_{o_2 o_1 \bar m,o_4 o_3 \bar n}^\Omega  ({\bf{q}}),  \hspace{1.5pc}\\
C(\text{or } D)_{o_1 o_2 m,o_3 o_4 n}^\Omega  ({\bf{q}}) = & e^{i{\bf{q}} \cdot ({\bf{R}}_n  - {\bf{R}}_m )} \\
& \times C(\text{or } D)_{o_4 o_3 \bar n,o_2 o_1 \bar m}^\Omega  ( - {\bf{q}})
\end{split}
\end{eqnarray}
for the remnant of antisymmetry of Grassmann variables. In Eq. (\ref{eq19}) the index $\bar m$ is associated with the Bravais lattice vector $-{\bf{R}}_m$.

\begin{figure}
	\begin{center}
		\includegraphics[width=8.6cm]{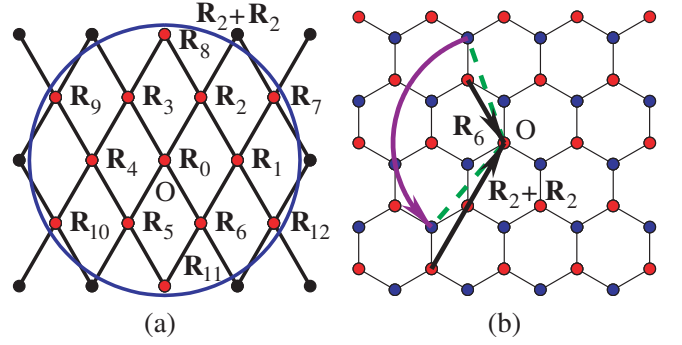}
	\end{center}
	\caption{(Color online) (a) Bravais lattice vectors ${\bf{R}}_0 \sim {\bf{R}}_{12}$ in the 13 plane-wave bases $f_m ({\bf{p}}) = e^{i{\bf{R}}_m \cdot {\bf{p}}}$ used by us. Only the bases with the Bravais lattice vectors (small red disks) inside the large blue circle are considered in our calculation. (b) Illustration of $2\pi /3$ rotation of a pair of atoms in the sublattice $A$ and $B$. Small red (blue) disks denote the sites in the sublattice $A$ ($B$). The index $(oo'm)$ represents a pair of atoms with one in the sublattice $o$ of the unit cell at the origin and other in the sublattice $o'$ of the unit cell at the position $-{\bf{R}}_m$. By $2\pi /3$ rotation, the atom in the sublattice $B$ of the unit cell at the positon $-{\bf{R}}_6$ is transferred to the one of the unit cell at the positon $-({\bf{R}}_2  + {\bf{R}}_2)$, which implies the symmetry relation between the indices $(A,B,{\bf{R}}_6)$ and $(A,B,{\bf{R}}_2 + {\bf{R}}_2 )$.}
	\label{fig1}
\end{figure}

In order for the relation (\ref{eq17}) to be exactly satisfied, some kind of {\em filtering process} is needed in each step of integration of the flow equations (\ref{eq14}). From Eq. (\ref{eq17}) we obtain
\begin{eqnarray} \label{eq20}
\begin{split}
& \left| {X_{\tilde o_1 ,\tilde o_2 ,Q{\bf{R}}_m  + {\bf{u}}_{o_1 }  - {\bf{u}}_{o_2 } ;\tilde o_3 ,\tilde o_4 ,Q{\bf{R}}_n  + {\bf{u}}_{o_3 }  - {\bf{u}}_{o_4 } }^\Omega  (Q{\bf{q}})} \right| \\
& \hspace{2pc} = \left| {X_{o_1 o_2 m,o_3 o_4 n}^\Omega  ({\bf{q}})} \right|
\end{split}
\end{eqnarray}
for the bosonic propagator $X$ ($X$ can be $P$, $C$, or $D$). From the point of view of numerical implementation, only a limited number of bases are involved in real calculation. In this work we use only 13 plane-wave bases with the Bravais lattice vectors shown in Fig. \ref{fig1}(a).

To be specific, we take $2\pi /3$ rotation about the origin as an example. It is easy to see that, in this case, the vector ${\bf{u}}_o$ and the sublattice $\tilde o$ are
\begin{equation}
\nonumber
{\bf{u}}_A  = {\bf{R}}_0 ,{\bf{u}}_B  = {\bf{R}}_5 ;\tilde A = A,\tilde B = B.
\end{equation}
According to Eq. (\ref{eq20}), the row (column) index $(AB6) = (A,B,{\bf{R}}_6)$ is related to the row (column) index $(\tilde A,\tilde B,Q{\bf{R}}_6  + {\bf{u}}_A  - {\bf{u}}_B ) = (A,B,{\bf{R}}_2  + {\bf{R}}_0  - {\bf{R}}_5 ) = (A,B,{\bf{R}}_2  + {\bf{R}}_2 )$, as shown in Fig. \ref{fig1}(b):
\begin{equation}
\nonumber
(A,B,{\bf{R}}_6 ) \leftrightarrow (A,B,{\bf{R}}_2  + {\bf{R}}_2 )
\end{equation}
Since the vector ${\bf{R}}_2  + {\bf{R}}_2$ is outside of the region of our consideration [see Fig. \ref{fig1}(a)], all the matrix elements with the row or column index of $(o,o',{\bf{R}}_2  + {\bf{R}}_2)$ are approximated to zero. This means that, according to Eq. (\ref{eq20}), the matrix elements $X_{AB6,oo'm}^\Omega  ({\bf{q}})$ and $X_{oo'm,AB6}^\Omega ({\bf{q}})$ have to vanish for all $(oo'm)$ and $\bf{q}$. Similarly, if the index $(oo'm)$ is related, under any symmetry operation, to the index $(\tilde o \tilde o' \tilde m)$ with ${\bf{R}}_{\tilde m}$ outside of the region of our consideration, then all the matrix elements in the row and the column associated with the index $(oo'm)$ have to be eliminated.

Table \ref{tab01} shows the row or column indices $(oo'm) = (o,o',{\bf{R}}_m )$ that have to be eliminated by the symmetry relation in our filtering process. Furthermore, this constraint should also be applied to the projection matrices $V^{\rm{P}} (\Omega)$, $V^{\rm{C}} (\Omega)$, and $V^{\rm{D}} (\Omega)$. Thus, in each step of integration of the flow equations (\ref{eq14}), the matrices $P^\Omega , C^\Omega  , D^\Omega  , V^{\rm{P}} (\Omega ), V^{\rm{C}} (\Omega)$, and $V^{\rm{D}} (\Omega)$ should be {\em filtered}, namely, all the elements with the row or column indices shown in Table \ref{tab01} have to be set to zero.

\begin{table}
	\caption{The row or column indices $(oo'm) = (o,o',{\bf{R}}_m )$ that should be eliminated in the filtering process.}
	\begin{center}
		\begin{tabular}{|c|c|}
			\hline
			 Sublattice indices & Bravais lattice vectors ${\bf{R}}_m$\\ 
			 $o,o'$  & of the plane-wave bases\\ 
			\hline
			$A,B$ & ${\bf{R}}_5,{\bf{R}}_6,{\bf{R}}_7,{\bf{R}}_9 ,{\bf{R}}_{10},{\bf{R}}_{11},{\bf{R}}_{12}$\\
			$B,A$ & ${\bf{R}}_2,{\bf{R}}_3,{\bf{R}}_7,{\bf{R}}_8 ,{\bf{R}}_9,{\bf{R}}_{10},{\bf{R}}_{12}$\\
			\hline
		\end{tabular}
	\end{center}
	\label{tab01}
\end{table}

On the other hand, the TUFRG approach provides an unbiased analysis of possible many-body instabilities in interacting electron systems. Several methods for determining the leading instabilities and corresponding order parameters are suggested. In many FRG studies addressing multiband systems, the final effective interaction at a critical scale $\Omega _C$ has been plugged into the mean-field equations. Wang {\it et al.} \cite{ref48} proposed an efficient FRG+MF procedure for computing order parameters in the systems with competing instabilities, in which only the irreducible part of the effective interaction entered the mean-field equations. Some of the authors have proposed a linear-response-based approach for identifying the type of order \cite{ref35}, which has been applied to the half-filled honeycomb lattice. The approach supports a high speed estimation of the form factors of the order parameters, so we will use it in the present work to determine the leading instabilities and build the phase diagrams. In the following, we briefly outline the approach.

One can identify the leading instability of the system by introducing infinitesimal test fields that are coupled to the fermion bilinears corresponding to various types of order and have the strength of $\lambda$. The form factors of the order parameters can be determined by considering the linear responses of the system to the test fields. For the singlet pairing, the triplet pairing, the spin and the charge channels, the order parameters in real space are defined as follows, respectively:
\begin{eqnarray} \label{eq21}
\begin{split}
& \Pi _{oo'}^{{\rm{sSC}}} ({\bf{R}}_i ,{\bf{R}}_i  - {\bf{R}}_\alpha  ) \\
& \hspace{2pc} = \mathop {\lim }\limits_{\lambda  \to  + 0} \sum\limits_\sigma  {\sigma \left\langle {c_{{\bf{R}}_i ,o,\sigma }^ \dag  c_{{\bf{R}}_i  - {\bf{R}}_\alpha  ,o', - \sigma }^ \dag  } \right\rangle _\lambda}, \\
& \Pi _{oo'}^{{\rm{tSC}}} ({\bf{R}}_i ,{\bf{R}}_i  - {\bf{R}}_\alpha  ) \\
& \hspace{2pc} = \mathop {\lim }\limits_{\lambda  \to  + 0} \sum\limits_\sigma  {\left\langle {c_{{\bf{R}}_i ,o,\sigma }^ \dag  c_{{\bf{R}}_i  - {\bf{R}}_\alpha  ,o', - \sigma }^ \dag  } \right\rangle _\lambda  }, \\
& \Pi _{oo'}^{{\rm{SPN}}} ({\bf{R}}_i ,{\bf{R}}_i  - {\bf{R}}_\alpha  ) \\
& \hspace{2pc} = \mathop {\lim }\limits_{\lambda  \to  + 0} \sum\limits_\sigma  \sigma  \left\langle {c_{{\bf{R}}_i ,o,\sigma }^ \dag  c_{{\bf{R}}_i  - {\bf{R}}_\alpha ,o',\sigma } } \right\rangle _\lambda, \\
& \Pi _{oo'}^{{\rm{CHG}}} ({\bf{R}}_i ,{\bf{R}}_i  - {\bf{R}}_\alpha  ) \\
& \hspace{2pc} = \mathop {\lim }\limits_{\lambda  \to  + 0} \sum\limits_\sigma  {\left\langle {c_{{\bf{R}}_i ,o,\sigma }^ \dag  c_{{\bf{R}}_i  - {\bf{R}}_\alpha  ,o',\sigma } } \right\rangle _\lambda}.
\end{split}
\end{eqnarray}
Here $\left\langle  \cdot  \right\rangle _\lambda$ means the ensemble average in the presence of corresponding test fields with coupling strength $\lambda$. All the order parameters above, except for the charge channel, vanish in the system without any spontaneous symmetry breaking. For the charge channel it exhibits the same symmetry as the system. However, if the system approaches a critical point, the corresponding susceptibility diverges and the order parameter could take a finite value. The fluctuation-dissipation theorem tells us that the divergent susceptibility leads to the divergence of related correlation function, which is, in the TUFRG calculation, represented by the divergence of the effective interaction and the bosonic propagator in corresponding channel. By Taylor expanding the ensemble averages in Eq. (\ref{eq21}) with respect to $\lambda$ and analyzing their divergence, we can find the expressions for the order parameters and identify the type of order. If only one type of order with a momentum transfer $\bf{Q}$ emerges in the system, the order parameter is expressed as \cite{ref35}
\begin{eqnarray} \label{eq22}
\begin{split}
\Pi _{oo'}^{\rm{X}} ({\bf{R}}_i ,{\bf{R}}_i  - {\bf{R}}_\alpha  ) = & Ce^{ - i{\bf{Q}} \cdot {\bf{R}}_i } \left[ {\phi _{oo'\alpha }^1 ({\bf{Q}})} \right]^* \\
\textrm {for } & \textrm{X=sSC or tSC,} \\
\Pi _{oo'}^{\rm{X}} ({\bf{R}}_i ,{\bf{R}}_i  - {\bf{R}}_\alpha  ) = & Ce^{ - i{\bf{Q}} \cdot {\bf{R}}_i } \left[ {\phi _{oo'\alpha }^1 ({\bf{Q}})} \right]^*  \\
& + C^* e^{i{\bf{Q}} \cdot ({\bf{R}}_i  - {\bf{R}}_\alpha  )} \phi _{o'o\bar \alpha }^1 ({\bf{Q}}) \\
\textrm {for } & \textrm{X=SPN or CHG.} \\
\end{split}
\end{eqnarray}
Here the constants $\phi _{oo'\alpha }^1 ({\bf{Q}})$ are the elements of the eigenvectors, associated with the most positive eigenvalues, of the following matrices:
\begin{eqnarray} \label{eq23}
\begin{split}
W^{{\rm{sSC}}} ({\bf{Q}}) & = W^{{\rm{tSC}}} ({\bf{Q}}) = \chi ^{{\rm{pp}}} ({\bf{Q}})[ - V^{\rm{P}} ({\bf{Q}})]\chi ^{{\rm{pp}}} ({\bf{Q}}), \hspace{1.5pc} \\
W^{{\rm{SPN}}} ({\bf{Q}}) & = \chi ^{{\rm{ph}}} ({\bf{Q}})V^{\rm{C}} ({\bf{Q}})\chi ^{{\rm{ph}}} ({\bf{Q}}), \\
W^{{\rm{CHG}}} ({\bf{Q}}) & = \chi ^{{\rm{ph}}} ({\bf{Q}})[V^{\rm{C}} ({\bf{Q}}) - 2V^{\rm{D}} ({\bf{Q}})]\chi ^{{\rm{ph}}} ({\bf{Q}}),
\end{split}
\end{eqnarray}
with the particle-particle and particle-hole susceptibility matrices,
\begin{eqnarray} \label{eq24}
\begin{split}
& \chi _{o'_1 o'_2 m,o_1 o_2 n}^{{\rm{pp}}} ({\bf{q}}) =  - \frac{1}{{S_{BZ} }}\int {d{\bf{k}}} f_m ({\bf{k}})f_n^* ({\bf{k}}) \\
& \hspace{2pc} \times \left[ {\frac{1}{\beta }\sum\limits_\omega  {G_{o'_1 o_1 }^0 } (\omega ,{\bf{k}} + {\bf{q}})G_{o'_2 o_2 }^0 ( - \omega , - {\bf{k}})} \right], \hspace{2pc} \\
& \chi _{o'_1 o'_2 m,o_1 o_2 n}^{{\rm{ph}}} ({\bf{q}}) =  - \frac{1}{{S_{BZ} }}\int {d{\bf{k}}} f_m ({\bf{k}})f_n^* ({\bf{k}}) \\
& \hspace{2pc} \times \left[ {\frac{1}{\beta }\sum\limits_\omega  {G_{o'_1 o_1 }^0 } (\omega ,{\bf{k}} + {\bf{q}})G_{o_2 o'_2 }^0 (\omega ,{\bf{k}})} \right].
\end{split}
\end{eqnarray}

For numeric implementation we discretize the irreducible region of the BZ by sampling points. The more sampling points involved, the more reliable results are expected, but the more computational effort is needed. This is true for the form-factor truncation. As mentioned above, we use 13 form-factor bases in TUFRG calculation. If we increase the number of bases to 19 (up to third intrasublattice nearest neighbors), the truncation error would be reduced, however, the required CPU time would become more than twice longer. So it is important to control the balance between both the reliability of the result and the computational effort. The choice of 13 bases truncation is justified by the fact that, in this case, the bare interaction is projected exactly onto three channels without any loss and the orbital picture used in this work ensures much faster convergence than the band picture in an expansion of the single-channel coupling functions in the bosonic propagators. This truncation has also been applied in previous work based on the band picture of TUFRG \cite{ref43}. For discretizing the BZ, we use a mesh of the momentum transfers of $N_{\bf{q}}=74$ points for the particle-particle channel and a mesh of $N_{\bf{q}}=98$ points for the particle-hole channel, as described in the following section. When Fourier transforming the bosonic propagators [it is needed to calculate the crossed contributions to the projection matrices, see Eqs. (\ref{eqC8})$-$(\ref{eqC11}) in Appendix \ref{appendC}], we introduce a linear fitting of the propagators in each triangle segments of the BZ while taking exactly the exponential functions to integrate analytically, which improves the quality of calculation. The resulting instabilities are robust with respect to further inclusion of the bases or introducing the denser meshes. The convergence tests for several points in parameter space, with truncation up to third neighbors or with doubly increased mesh points, have shown moderate variations in the resultant critical scales.

\section{Results and Discussion}\label{sec3}

The projection matrices $V^{\rm{P}}(\Omega)$, $V^{\rm{C}}(\Omega)$ and $V^{\rm{D}}(\Omega)$ enter the flow equations for the bosonic propagators, so they should be found to integrate out the flow equation. Since the effective interaction is represented via the bosonic propagators,
\begin{eqnarray} \label{eq25}
\begin{split}
V^\Omega & = V^{(0)}+\Phi^{{\rm{pp}}} (\Omega ) + \Phi ^{{\rm{ph,cr}}} (\Omega ) + \Phi ^{{\rm{ph,d}}} (\Omega) \\
& \approx V^{(0)}+{\rm{\hat P}}^{-1} [P^\Omega]+{\rm{\hat C}}^{-1} [C^\Omega] + {\rm{\hat D}}^{-1} [D^\Omega], \hspace{2pc}
\end{split}
\end{eqnarray}
the projection matrices can also be expressed in terms of $P^\Omega$, $C^\Omega$ and $D^\Omega$ [the detailed expressions for the projection matrices are given in Eqs. (\ref{eqC8})$-$(\ref{eqC10}) in Appendix \ref{appendC}]:
\begin{eqnarray} \label{eq26}
\begin{split}
V^{\rm{P}} (\Omega ) \approx & {\rm{\hat P}}[V^{(0)} ] + P^\Omega \\
& + {\rm{\hat P\{ \hat C}}^{ - 1} [C^\Omega  ]\}  + {\rm{\hat P\{ \hat D}}^{ - 1} [D^\Omega  ]\} , \\ 
V^{\rm{C}} (\Omega ) \approx & {\rm{\hat C}}[V^{(0)} ] + C^\Omega \\
& + {\rm{\hat C\{ \hat P}}^{ - 1} [P^\Omega  ]\}  + {\rm{\hat C\{ \hat D}}^{ - 1} [D^\Omega  ]\} , \\ 
V^{\rm{D}} (\Omega ) \approx & {\rm{\hat D}}[V^{(0)} ] + D^\Omega \\
& + {\rm{\hat D\{ \hat P}}^{ - 1} [P^\Omega  ]\}  + {\rm{\hat D\{ \hat C}}^{ - 1} [C^\Omega  ]\}.
\end{split}
\end{eqnarray}
By substituting this into Eq. (\ref{eq14}) we obtain a closed system of differential equations for the matrices $P^\Omega$, $C^\Omega$ and $D^\Omega$. These matrices have vanishing initial values, and the initial values of $V^{\rm{P}}(\Omega)$, $V^{\rm{C}}(\Omega)$ and $V^{\rm{D}}(\Omega)$, namely, $V^{{\rm{P}},(0)} \equiv {\rm{\hat P}}[V^{(0)}]$, $V^{{\rm{C}},(0)} \equiv {\rm{\hat C}}[V^{(0)}]$ and $V^{{\rm{D}},(0)} \equiv {\rm{\hat D}}[V^{(0)}]$ are needed for solving the system of equations. These are determined by Fourier transforming Eq. (\ref{eq02}) and projecting it onto three channels. The results are as follows:
\begin{eqnarray} \label{eq27}
\begin{split}
V_{AA0,AA0}^{{\rm{P}}({\rm{C}},{\rm{D}}),(0)} ({\bf{q}}) &= V_{BB0,BB0}^{{\rm{P}}({\rm{C}},{\rm{D}}),(0)} ({\bf{q}}) = U,\\
V_{ABm,ABm}^{{\rm{P}}({\rm{C}}),(0)} ({\bf{q}}) &= V \hspace{1pc} (m = 0,2,3),\\
V_{BAm,BAm}^{{\rm{P}}({\rm{C}}),(0)} ({\bf{q}}) &= V \hspace{1pc} (m = 0,5,6),\\
V_{ABm,ABm}^{{\rm{D}},(0)} ({\bf{q}}) &= J \hspace{1pc} (m = 0,2,3),\\
V_{BAm,BAm}^{{\rm{D}},(0)} ({\bf{q}}) &= J \hspace{1pc} (m = 0,5,6),\\
V_{AB0,BA0}^{{\rm{P}}({\rm{C}},{\rm{D}}),(0)} ({\bf{q}}) &= V_{BA0,AB0}^{{\rm{P}}({\rm{C}},{\rm{D}}),(0)} ({\bf{q}}) = J,\\
V_{ABm,BA\bar m}^{{\rm{P}}({\rm{C}},{\rm{D}}),(0)} ({\bf{q}}) &= \left[ {V_{BA\bar m,ABm}^{{\rm{P}}({\rm{C}},{\rm{D}}),(0)} ({\bf{q}})} \right]^*  \\
&= Je^{ - i{\bf{R}}_m  \cdot {\bf{q}}} \hspace{1pc} (m = 2,3),\\
V_{AA0,BB0}^{{\rm{P}}({\rm{C}}),(0)} ({\bf{q}}) &= \left[ {V_{BB0,AA0}^{{\rm{P}}({\rm{C}}),(0)} ({\bf{q}})} \right]^* \\
&= J(1 + e^{ - i{\bf{R}}_2  \cdot {\bf{q}}}  + e^{-i{\bf{R}}_3  \cdot {\bf{q}}}),\\
V_{AA0,BB0}^{{\rm{D}},(0)} ({\bf{q}}) &= \left[ {V_{BB0,AA0}^{{\rm{D}},(0)} ({\bf{q}})} \right]^* \\
&= V(1 + e^{ - i{\bf{R}}_2  \cdot {\bf{q}}}  + e^{ - i{\bf{R}}_3  \cdot {\bf{q}}}),\\
\text{All other elements} &= 0.
\end{split}
\end{eqnarray}

\begin{figure}[h!]
	\begin{center}
		\includegraphics[width=8.6cm]{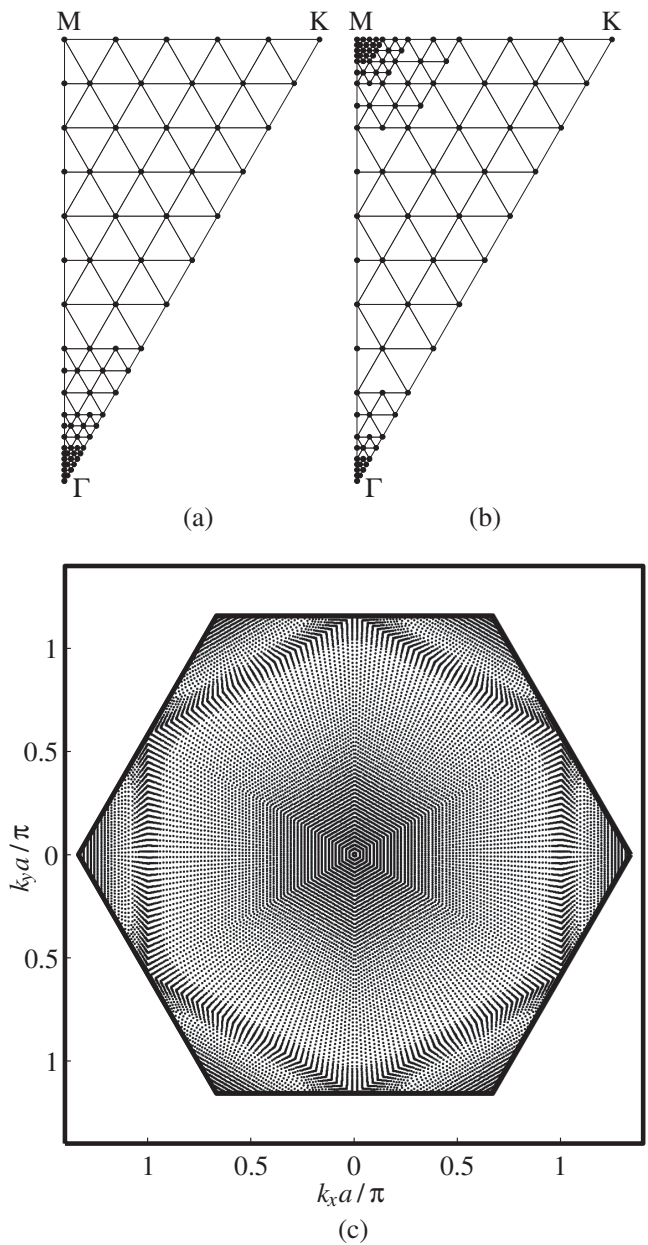}
	\end{center}
	\caption{(a) Mesh of $N_{\bf{q}}=74$ points for momentum transfers within the irreducible region of BZ in the particle-particle channel. The points are distributed more densely near the $\bf{\Gamma}$ point. The bosonic propagator $P(\bf{q})$ is calculated for these points. (b) Mesh of $N_{\bf{q}}=98$ points for momentum transfers within the irreducible region of BZ in the particle-hole channel. The points are distributed more densely near the $\bf{\Gamma}$ and $\bf{M}$ points. The bosonic propagators $C(\bf{q})$ and $D(\bf{q})$ are calculated for these points. (c) Mesh of $N_{\bf{k}}=40320$ points for sampling momenta used in the integration of $\dot \chi^{{\rm{pp}}}$ and $\dot \chi^{{\rm{ph}}}$ for the doping level at the VHS ($\delta=0.25$). Here the points are distributed more densely near the FS, while $a \approx 2.46 {\rm{\AA}}$ is the lattice constant, namely, the distance between next-nearest-neighbor sites.}
	\label{fig2}
\end{figure}

In our calculation the matrices $P^\Omega$, $C^\Omega$, $D^\Omega$, $V^{\rm{P}}(\Omega)$, $V^{\rm{C}}(\Omega)$, $V^{\rm{D}}(\Omega)$, $\chi^{{\rm{pp}}}$, and $\chi^{{\rm{ph}}}$ have $N_D \times N_D$ structures with $N_D=2 \times 2 \times 13-14=38$ reduced via the filtering process. The flow equations for the bosonic propagators, Eq. (\ref{eq14}), are solved only for the momentum transfers in the irreducible region of the BZ which are shown in Fig. \ref{fig2}(a) and \ref{fig2}(b). The mesh of the momentum transfers ($\bf{q}$ mesh) are constructed such that the discretized momentum transfers, i.e., $\bf{q}$ vectors, are distributed more densely near the high-symmetry points that are most likely candidates for the possible ordering vectors. In each step of integration of the equations, the bosonic propagators outside of the region are generated by the symmetry relations (\ref{eq17}) and then plugged into Eq. (\ref{eq26}) to produce the projection matrices. Figure \ref{fig2}(c) shows the sampling momenta used in the integration of $\dot \chi^{{\rm{pp}}}$ and $\dot \chi^{{\rm{ph}}}$ in Eq. (\ref{eq15}) for the doping level at the VHS ($\delta=0.25$), which are denser near the FS.

\begin{figure*}
	\begin{center}
		\includegraphics[width=17.2cm]{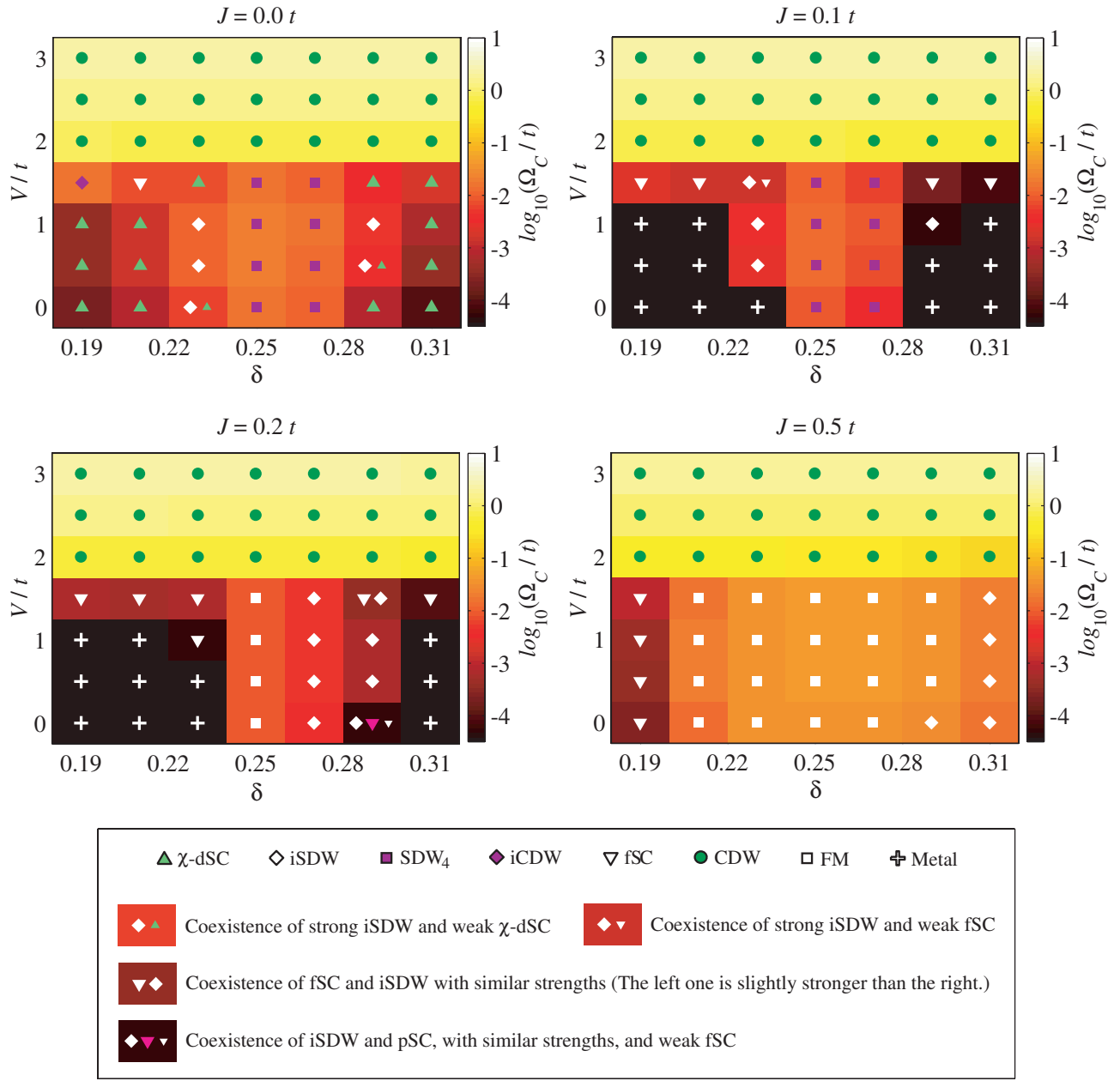}
	\end{center}
	\caption{(Color online) Schematic phase diagrams for small and moderate values of $J$. The color bars indicate values of critical scales $\Omega _C$ at which the corresponding transitions may occur. In the region denoted as Metal, there is no divergence of any bosonic propagator in the RG flow down to the stopping scale $\Omega ^*=1.3 \times 10^{-4} {\rm{eV}}$. In the coexistence regions, there are two or three dominant eigenmodes of several bosonic propagators. Here, different orders could coexist or exclude the others, or even may compete with each other leading to common suppression. The notation pSC is a shorthand notation for the $p$-wace SC.}
	\label{fig3}
\end{figure*}

\begin{figure}
	\begin{center}
		\includegraphics[width=8.6cm]{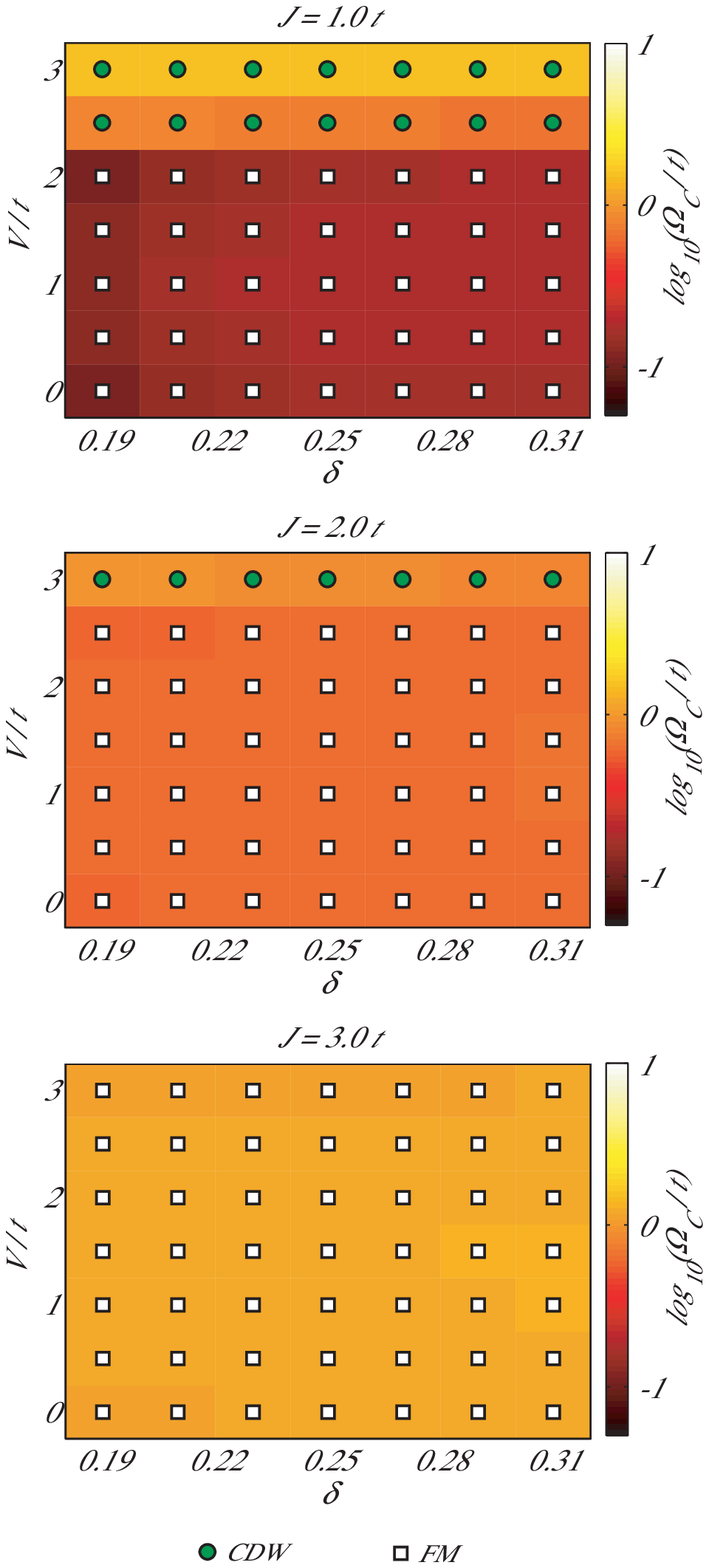}
	\end{center}
	\caption{(Color online) Schematic phase diagrams for large values of $J$. The color bars indicate values of critical scales $\Omega_C$. The phase diagrams have simple structures consisting only of two phases, namely, a $\pi /3$-rotation-symmetry-broken charge-density-wave (CDW) phase and a ferromagnetic (FM) phase.}
	\label{fig4}
\end{figure}

The ordering tendencies towards diverse symmetry-broken ground states are analyzed by means of the linear-response-based approach described in Sec.~\ref{sec2C}. We have investigated these tendencies by varying the parameters $\delta$ and $V$, while fixing $J$ and $U$. The results for small and moderate values of $J$ are summarized in tentative phase diagrams shown in Fig. \ref{fig3}, while those for large values of $J$ in Fig. \ref{fig4}. The critical scales $\Omega _C$, at which a divergence of corresponding bosonic propagator is observed, are also provided using the color bars. We outline some features of the phase diagrams below.\\

In the absence of the ferromagnetic exchange ($J = 0.0t$) and for small values of the nearest-neighbor repulsion ($V = 0 \sim t$), a four-sublattice spin-density-wave (SDW$_4$) phase occurs at and close to the VHS filling, while the chiral $d$-wave superconducting ($\chi$-dSC) phase away from it. An incommensurate spin-density-wave (iSDW) phase occupies the regions between the SDW$_4$ and the $\chi$-dSC. This configuration is very similar to that in Ref. \onlinecite {ref26}. For moderate values of $V$ ($V \approx 1.5t$), the SDW$_4$ is found again in the vicinity of the VHS, while two $\chi$-dSC regions flank it. Two additional phases, namely, a spin-triplet $f$-wave superconducting (fSC) and an incommensurate charge-density-wave (iCDW) phases occur at lower doping levels ($\delta  \approx 0.19$ for iCDW and $\delta  \approx 0.21$ for fSC). For large values of $V$ ($V = 2t \sim 3t$), a charge-density-wave phase with broken $\pi /3$-rotation symmetry and a charge transfer from sublattice $A$ to $B$ (or vice versa), which is denoted as CDW in Figs. \ref{fig3} and \ref{fig4}, is found for all doping levels considered. The associated critical scales are considerably increased.

When a weak exchange coupling ($J=0.1t$) is involved, the structure of the phase diagram exhibits a remarkable change. The $\chi$-dSC phase, found for $J=0, V=0 \sim 1.5t$, completely disappears from our phase diagram. The phase turns into a metallic phase for $V=0 \sim t$, while into the fSC for $V \approx 1.5t,\delta  = 0.29 \sim 0.31$ and a coexistence phase of strong iSDW and weak fSC orders for $V \approx 1.5t,\delta  \approx 0.23$. In the coexistence phase, the two orders, i.e., the iSDW and fSC could coexist or exclude the other, or even may compete with each other leading to common suppression. Since we have not performed the mean-field calculation, we cannot determine whether these orders really coexist or not, as well as, the relative strength of corresponding order parameters, if they coexist \footnote{The coexistence phases in Fig. \ref{fig3} have not been rigorously verified. They were identified by comparing the strengths of divergences for dominant eigenmodes of the $W$ matrices in Eq. (\ref{eq23}). Concretely, the notation \emph{Coexistence of strong A and weak B} means that, at the critical scale, the eigenvalue of $W$ matrix, associated with the A phase, $\lambda_{\rm{A}}$ is $2 \sim 5$ times larger than that associated with the B phase, $\lambda_{\rm{B}}$ (i.e., $\lambda_{\rm{A}}/5 < \lambda_{\rm{B}}  \le \lambda_{\rm{A}}/2$). The notation \emph{Coexistence of A and B with similar strengths} means the relation $\lambda_{\rm{A}}/2 < \lambda_{\rm{B}}  \le \lambda_{\rm{A}}$ at the critical scale, while \emph{Coexistence of A and B, with similar strengths, and weak C} means $\lambda_{\rm{A}}/2 < \lambda_{\rm{B}}  \le \lambda_{\rm{A}}$ as well as $\lambda_{\rm{A}}/5 < \lambda_{\rm{C}}  \le \lambda_{\rm{A}}/2$.}. The iCDW at $V \approx 1.5t,\delta  \approx 0.19$ changes to the fSC. The whole SDW$_4$ region and main part of the iSDW region survive with decreased critical scales. The fSC at $V \approx 1.5t,\delta  \approx 0.21$ and the CDW for $V = 2t\sim 3t$ are also retained.

It is very interesting that the metallic phase is induced by involving exchange coupling. This behavior can be attributed to competition effect, as evidenced by the suppression of critical scales near the boundaries between different phases. The present TUFRG scheme is apparently more sensitive to competition effects due to its high momentum resolution. Although the current version of TUFRG, as an approach from the weak-coupling perspective, is not certainly exact and is on its development, the physically plausible observation of a metallic state being stabilized by competition effects deserves to be considered thoroughly and compared with other methods. This interesting result has also been obtained in previous TUFRG studies on half-filled honeycomb lattice \cite{ref35, ref43}, in which a semimetallic state was recovered by increasing some interaction parameters.

If the exchange coupling is further increased ($J=0.2t$), then the iSDW for $V = 0.5t\sim 1.5t,\delta  \approx 0.23$ turns into the fSC, while it develops in the region of $V = 0 \sim 1.5t,\delta  \approx 0.29$. The SDW$_4$ changes to a ferromagnetic (FM) phase for $\delta  \approx 0.25$ and the iSDW for $\delta  \approx 0.27$. In the case of relatively large value of $J$ ($J = 0.5t$), the FM phase occupies a large part of the region for $V = 0\sim 1.5t$. Both sides of the region are occupied by the fSC (left) and the iSDW (right). The CDW region remains unchanged in the case of $J = 0\sim 0.5t$.

When the ferromagnetic exchange interaction is further enhanced, the phase diagrams have simple structures consisting only of two phases, i.e., the CDW and FM phases. For $J = t$, a transition from the FM to the CDW is found at $V \approx 2.25t$, independently on doping level. The transition point is moved to $V \approx 2.75t$ for $J=2t$, and finally, for $J=3t$, the whole parameter space of the phase diagram is covered by the FM phase.

From Fig. \ref{fig3}, we see that the chiral $d$-wave SC is destroyed upon increasing $J$ while the $f$-wave SC is developed. This can be explained by the Kohn-Luttinger mechanism \cite{ref55} in which the fluctuations in particle-hole channel (mostly spin channel) have cross contribution to particle-particle channel, finally resulting in attractive interaction in some paring channel. More specifically, in our case, the spin channel develops gradually in the high-energy scale followed by paring channel (mainly $s$-wave). In the intermediate RG stage the incommensurate SDW, with momentum transfer equal to the near-nesting vector of the FS geometry, develops fast and dominantly, which prevents the growth of $s$-wave but promotes attractive $d$-wave and $f$-wave. In the absence of exchange coupling, the $d$-wave channel grows fast and construct a positive mutual feedback with the incommensurate SDW at the lower-energy scale. At the low-energy scale the $d$-wave SC flows faster and diverges eventually at the critical scale. However, if the exchange coupling is involved, it is supposed that the $d$-wave channel gets additional repulsive interaction while the $f$-wave gets attractive. This behavior is originated from the tendency of ferromagnetic exchange to align spins in favor of spin-triplet paring. So in the intermediate stage the $f$-wave develops faster than $d$-wave and then builds a negative feedback with the SDW which suppresses the growths of both the $f$-wave and SDW instabilities. But larger nearest-neighbor repulsion also provides the $f$-wave with attractive interaction, so that, the $f$-wave can grow fast and diverge before making the negative feedback. Of course, this scenario has to be verified further by more detailed analysis and comparison with other results. Now, we describe concretely some exotic phases among those mentioned above.\\

{\it Chiral d-wave superconducting phase} ($\chi$-dSC)

\begin{figure}
	\begin{center}
		\includegraphics[width=8.6cm]{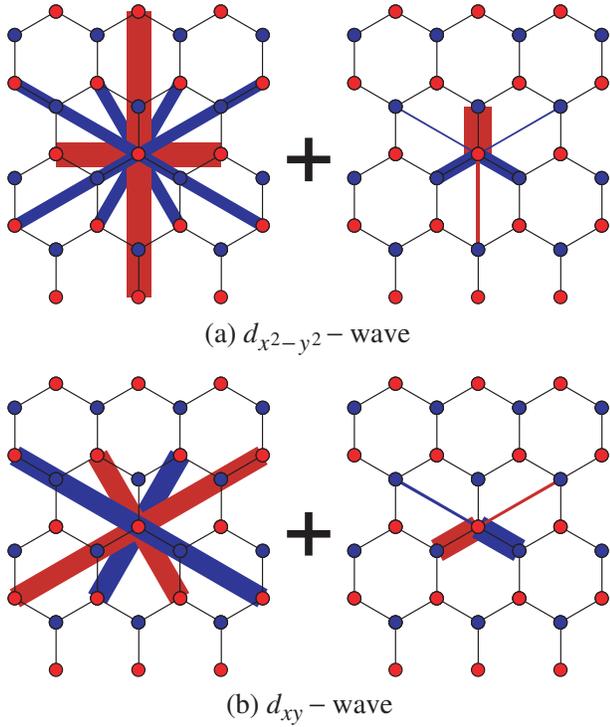}
	\end{center}
	\caption{(Color online) Form factors of the spin-singlet SC order parameters, $\Pi _{Ao}^{{\rm{sSC}}} (0,{\bf{R}}_m )$ ($o = A,B$), for $J = 0,V = 0.5t,\delta  = 0.21$ in the $d_{x^2  - y^2 }$-wave (a) and the $d_{xy}$-wave (b) states. The red (blue) sticks indicate the positive (negative) values of the SC order parameters, while the widths of sticks measure magnitudes of the order parameters. The order parameters $\Pi _{BB}^{{\rm{sSC}}} (0,{\bf{R}}_m )$ have the same form factors as $\Pi _{AA}^{{\rm{sSC}}} (0,{\bf{R}}_m)$, and $\Pi _{BA}^{{\rm{sSC}}} (0,{\bf{R}}_m)$ can be obtained using the relation $\Pi _{BA}^{{\rm{sSC}}} (0,{\bf{R}}_m ) = \Pi _{AB}^{{\rm{sSC}}} (0, - {\bf{R}}_m)$. The form factors for these two states can make a linear combination of $d_{x^2  - y^2 } \pm id_{xy}$ to form the chiral $d$-wave SC.}
	\label{fig5}
\end{figure}

\begin{figure}
	\begin{center}
		\includegraphics[width=8.6cm]{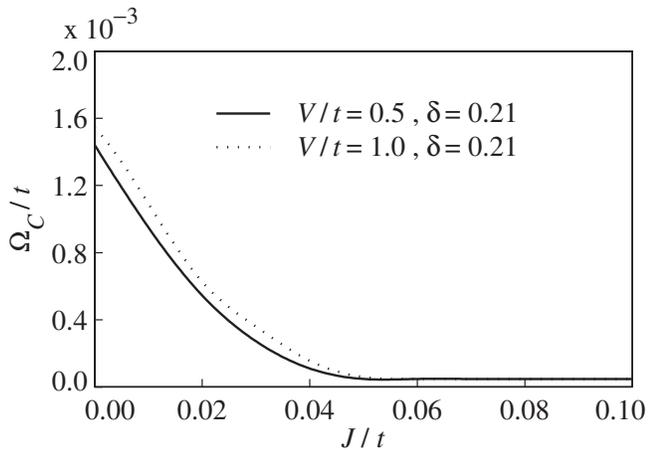}
	\end{center}
	\caption{Critical scales $\Omega _C$ of the SC transitions as function of exchange coupling $J$. The plots are given for $V = 0.5t,\delta  = 0.21$ (solid) and $V = 1.0t,\delta  = 0.21$ (dotted). The SC order is suppressed at $J \approx 0.05t$.}
	\label{fig6}
\end{figure}

The chiral $d$-wave SC phase manifests itself in the RG flow as two dominantly divergent and degenerate eigenmodes of $W^{{\rm{sSC}}} ({\bf{Q}} = 0)$ that obey the two-dimensional $E_2$ representation of $C_{6v}$ symmetry \cite{ref12, ref40}. These two modes can make a complex linear combination to form the chiral $d$-wave SC. In order to determine whether it is really formed or not, one needs to perform the mean-field calculation using the effective interaction, but this is beyond the scope of the present work. Many previous works support a formation of the chiral SC. The form factors of two kinds of the SC order parameters, which are associated with those two dominant modes and have $d_{x^2-y^2 }$ and $d_{xy}$ symmetries, are depicted in Fig. \ref{fig5}.

What is most surprising is the annihilation of this intriguing order by a weak exchange coupling. As mentioned above, the chiral $d$-wave SC has completely disappeared from our phase diagram upon including the exchange coupling of $J=0.1t$. The investigation by a fine tuning of the parameter $J$ shows that the phase is fully suppressed by weak exchange coupling of $J \approx 0.05t = 0.14{\rm{eV}}$ (see Fig. \ref{fig6}). Although we are not able to give an exact value of $J$ for graphene, our rough estimation presented in Appendix \ref{appendA} shows that the above value of $J$ is inside of the range of the expectation value for the exchange coupling, namely $0.10 \sim 0.28$eV. Thus, our results demonstrate possible destruction of the chiral SC in single layer graphene, which is consistent with a failure in experimental effort to search for it in the system.

{\it Four-sublattice spin-density-wave phase} (SDW$_4$)

\begin{figure*}
	\begin{center}
		\includegraphics[width=17.2cm]{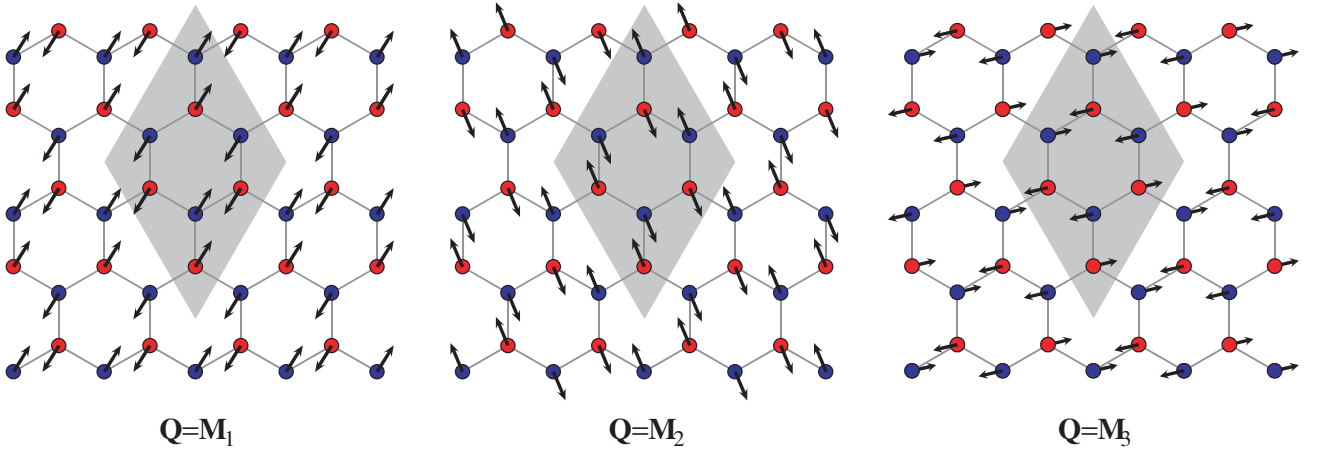}
	\end{center}
	\caption{(Color online) Spin distributions for the four-sublattice SDW phases with three inequivalent momentum transfers ${\bf{M}}_{1,2,3}$. The shaded area indicates the unit cell common to these spin patterns.}
	\label{fig7}
\end{figure*}

\begin{figure*}
	\begin{center}
		\includegraphics[width=17.2cm]{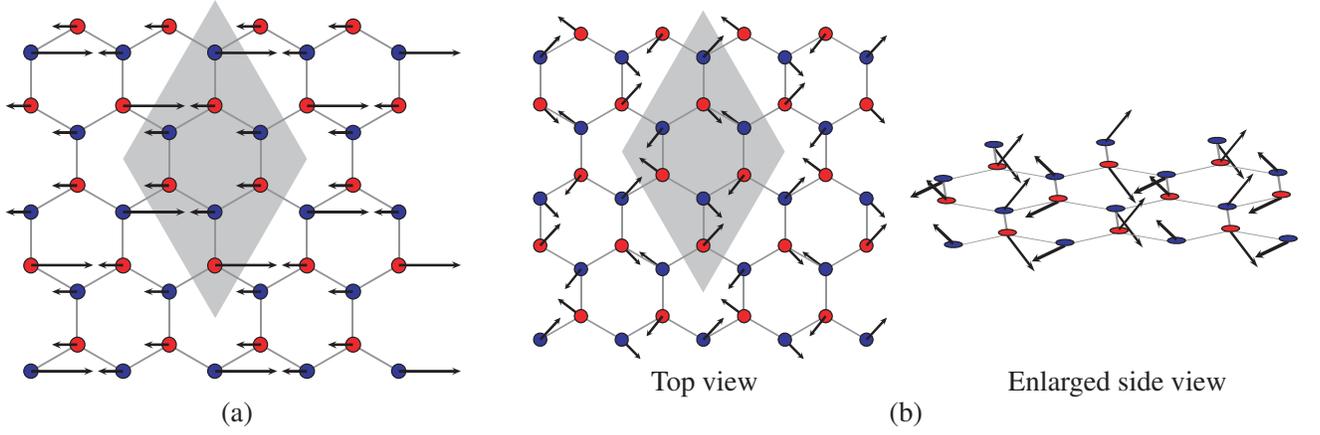}
	\end{center}
	\caption{(Color online) Spin distributions for the collinear SDW phase (a) and the chiral SDW phase (b). The shaded area indicates the unit cell of the phases.}
	\label{fig8}
\end{figure*}

The four-sublattice SDW phase manifests itself in our RG flow as a dominantly divergent eigenmode of $W^{{\rm{SPN}}} ({\bf{Q}} = {\bf{M}}_1)$ which has real numbers $\phi _{AA0}  = \phi _{BB0}$ as its largest components. It is driven by nearly perfect nesting of the FS. The momentum transfer ${\bf{M}}_1$ has two other partners, ${\bf{M}}_2$ and ${\bf{M}}_3$, to which it can be transformed by symmetry operations. The spin distributions for the phases with three inequivalent momentum transfers ${\bf{M}}_{1,2,3}$ are shown in Fig. \ref{fig7}. In general, the spin distribution in the system is realized by superposition of those patterns and can be represented by following equation:
\begin{eqnarray} \label{eq28}
\begin{split}
\left\langle {{\bf{\hat S}}_{Ai} } \right\rangle = & {\bf{S}}_1 \cos ({\bf{M}}_1  \cdot {\bf{R}}_i ) \\
& + {\bf{S}}_2 \cos ({\bf{M}}_2  \cdot {\bf{R}}_i ) + {\bf{S}}_3 \cos ({\bf{M}}_3  \cdot {\bf{R}}_i ), \\
\left\langle {{\bf{\hat S}}_{Bi} } \right\rangle = & {\bf{S}}_1 \cos ({\bf{M}}_1  \cdot {\bf{R}}_i ) \\
& - {\bf{S}}_2 \cos ({\bf{M}}_2  \cdot {\bf{R}}_i ) - {\bf{S}}_3 \cos ({\bf{M}}_3  \cdot {\bf{R}}_i ), \hspace{2pc}
\end{split}
\end{eqnarray}
with three nesting vectors,
\begin{equation}
\nonumber
\begin{split}
{\bf{M}}_1  = & \frac{{2\pi }}{{\sqrt 3 a}}(0,1), \\
{\bf{M}}_2  = & \frac{\pi }{{\sqrt 3 a}}( - \sqrt 3 , - 1), {\bf{M}}_3  = \frac{\pi }{{\sqrt 3 a}}(\sqrt 3 , - 1).
\end{split}
\end{equation}

Various patterns can be generated by arbitrary selection of three amplitudes ${\bf{S}}_1 \sim {\bf{S}}_3$. If one selects as ${\bf{S}}_1  = {\bf{S}}_2  = {\bf{S}}_3$, the collinear SDW phase emerges as suggested in Ref. \onlinecite {ref34} [see Fig. \ref{fig8}(a)]. The chiral SDW phase \cite{ref25, ref26, ref31, ref32} is generated by setting ${\bf{S}}_1 ,{\bf{S}}_2$ and ${\bf{S}}_3$ as three mutually orthogonal vectors [see Fig. \ref{fig8}(b)]. It is argued that the chiral SDW order is only developed at the lowest temperatures and turns into the collinear SDW phase at higher temperatures \cite{ref34}. Again, the mean-field calculation is needed to determine which of these phases is favored, but this is beyond the scope of this paper.

{\it Spin-triplet f-wave superconducting phase} (fSC)

\begin{figure}
	\begin{center}
		\includegraphics[width=8.6cm]{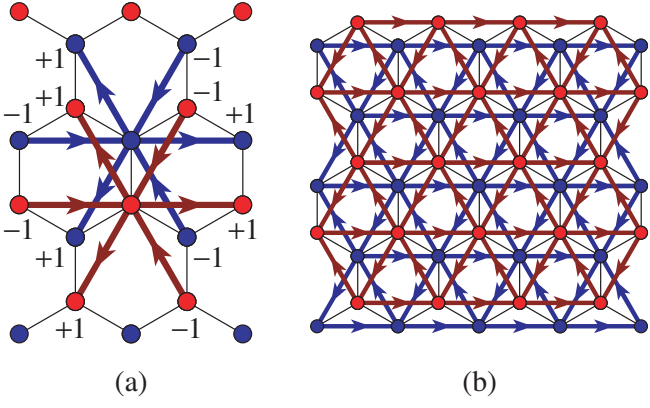}
	\end{center}
	\caption{(Color online) (a) Form factor of the spin-triplet SC order parameters, $\Pi _{AA}^{{\rm{tSC}}} (0,{\bf{R}}_m)$ and $\Pi _{BB}^{{\rm{tSC}}} (0,{\bf{R}}_m)$, for the $f$-wave SC phase. The outward (inward) arrows indicate positive (negative) values of the order parameters. (b) Spin-triplet SC order pattern in the phase, denoted by arrows. The arrow from the site $m$ to $n$ in the sublattice $o$ corresponds to the order parameter $\Pi _{oo}^{{\rm{tSC}}} ({\bf{R}}_m ,{\bf{R}}_n ) =  - \Pi _{oo}^{{\rm{tSC}}} ({\bf{R}}_n ,{\bf{R}}_m ) = 1$.}
	\label{fig9}
\end{figure}

The ferromagnetic exchange coupling and the nearest-neighbor repulsion favor the spin-triplet $f$-wave SC phase. It is represented by a predominant eigenmode of $W^{{\rm{tSC}}} ({\bf{Q}} = 0)$ that follows the one-dimensional $B_1$ representation of $C_{6v}$ symmetry \cite{ref23, ref40}. All the fSC phase in our phase diagrams have the same form factor of the SC order parameter, as shown in Fig. \ref{fig9}. This form factor is expressed as
\begin{equation}
\nonumber
\begin{split}
\Pi _{oo}^{{\rm{tSC}}} (0,{\bf{R}}_1 ) =& \Pi _{oo}^{{\rm{tSC}}} (0,{\bf{R}}_3 ) = \Pi _{oo}^{{\rm{tSC}}} (0,{\bf{R}}_5 ) = 1, \\
\Pi _{oo}^{{\rm{tSC}}} (0,{\bf{R}}_2 ) =& \Pi _{oo}^{{\rm{tSC}}} (0,{\bf{R}}_4 ) = \Pi _{oo}^{{\rm{tSC}}} (0,{\bf{R}}_6 ) =  - 1,
\end{split}
\end{equation}
which, by a Fourier transformation, presents the following order parameter in momentum space:
\begin{eqnarray} \label{eq29}
\begin{split}
\mathop {\lim }\limits_{\lambda  \to  + 0} \sum\limits_\sigma  {\left\langle {c_{{\bf{k}},o,\sigma }^ \dag  c_{ - {\bf{k}},o, - \sigma }^ \dag  } \right\rangle _\lambda  }  = \sum\limits_m {\Pi _{oo}^{{\rm{tSC}}} (0,{\bf{R}}_m )} e^{ - i{\bf{R}}_m  \cdot {\bf{k}}} \\
= - 2\left[ {\sin (k_x a) - 2\sin \left( {\frac{{k_x a}}{2}} \right)\cos \left( {\frac{{\sqrt 3 k_y a}}{2}} \right)} \right]. \hspace{2pc}
\end{split}
\end{eqnarray}
The above expression leads to nodal gap which has nodes on the nodal lines, $k_x  = 0$ and $k_x  =  \pm \sqrt 3 k_y$. For doping levels lower than the VHS filling, the FS is disconnected and the nodes of the gap do not intersect with the FSs. In this case the $f$-wave SC state could become fully gapped. In our calculation we have not found the $f$-wave gap that has been suggested in Ref. \onlinecite{ref23} and obeys the $B_2$ representation.

{\it Incommensurate spin-density-wave phase} (iSDW)

\begin{figure}
	\begin{center}
		\includegraphics[width=8.6cm]{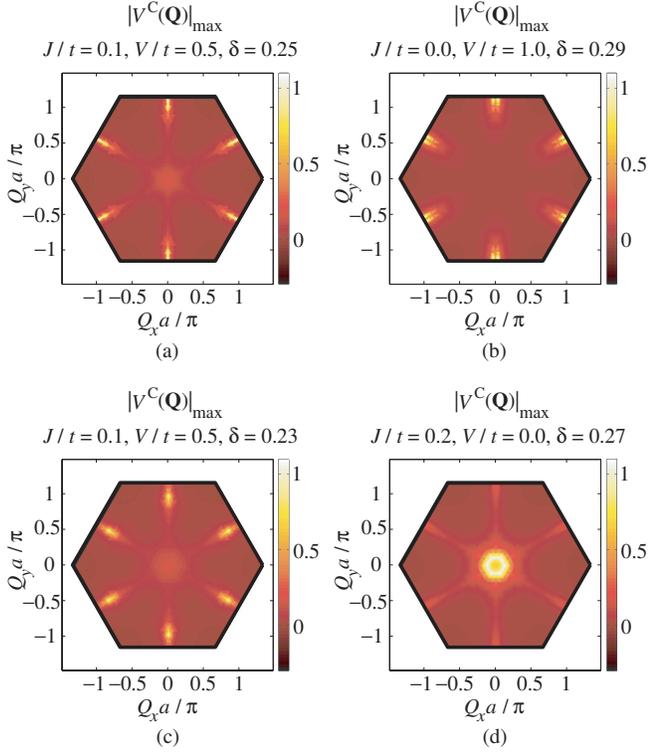}
	\end{center}
	\caption{(Color online) Maximum absolute values of the projection matrices $V^{\rm{C}} ({\bf{Q}})$ in the momentum space for the SDW$_4$ and iSDW phases. The color bars indicate the relative values $|V^{\rm{C}} ({\bf{Q}})|_{\max}/|V^{\rm{C}} ({\bf{Q_{\max}}})|_{\max}$. (a) Plot for $J = 0.1t,V = 0.5t,\delta  = 0.25$ (SDW$_4$). It has three peaks at the vectors ${\bf{M}}_1 ,{\bf{M}}_2$ and ${\bf{M}}_3$. (b) Plot for $J = 0,V = 1.0t,\delta  = 0.29$ (iSDW). (c) Plot for $J = 0.1t,V = 0.5t,\delta  = 0.23$ (iSDW). The plots (b) and (c) have six peaks near the vectors ${\bf{M}}_{1,2,3}$. (d) Plot for $J = 0.2t,V = 0,\delta  = 0.27$ (iSDW). It has six peaks around the ${\bf{\Gamma}}$ point.}
	\label{fig10}
\end{figure}

The incommensurate SDW phase manifests itself in the RG flow as a dominant eigenmode of $W^{{\rm{SPN}}} ({\bf{Q}} = {\bf{Q}}_0 )$ which has complex numbers $\phi _{AA0}$ and $\phi _{BB0}$ with the relation $\left| {\phi _{AA0} } \right| = \left| {\phi _{BB0} } \right|$ as its largest components. For several parameter sets within the SDW$_4$ and iSDW regions, the maximum absolute values of the projection matrix $V^{\rm{C}} ({\bf{Q}})$, denoted as $|V^{\rm{C}} ({\bf{Q}})|_{\max}$, are plotted as function of momentum transfer $\bf{Q}$ in Fig. \ref{fig10}. They have strong peaks at some ordering vectors which depend on the values of $\delta$ and $J$, but not on $V$. The plot has three peaks at the vectors ${\bf{M}}_{1,2,3}$ for the SDW$_4$ phase [see Fig. \ref{fig10}(a)]. When the parameter $J$ is small, the plot for the iSDW phase has six peaks near the vectors ${\bf{M}}_{1,2,3}$ [Figs. \ref{fig10}(b) and \ref{fig10}(c)]. The variation in the peak positions (ordering vectors) is related with the change in the FS shape. However, if $J$ is further increased, the peak positions are moved to around the ${\bf{\Gamma}}$ point [Fig. \ref{fig10}(d)] which is far from the nesting vectors of the near-nested FS. This noticeable change can be attributed to the competition between both ordering tendencies toward the ferromagnetic phase by increased $J$ and toward the SDW$_4$ due to near nesting of the FS.

\section{Conclusion} \label{sec4}
This work has addressed the effect of enhanced exchange interaction on the ground-state orderings of electrons on the honeycomb lattice doped to the vicinity of the VHS. An extended Hubbard model, including the on-site and nearest-neighbor Coulomb repulsions, and nearest-neighbor ferromagnetic exchange and pair hopping interactions, has been considered. The effective interactions have been calculated by using the TUFRG allowing for high momentum resolution, while the ground states of the system have been analyzed employing the linear-response-based approach for identifying the type of order. The ground-state phase diagrams in the space of doping level and nearest-neighbor repulsion were obtained for several values of nearest-neighbor exchange integral. Inclusion of small and moderate ferromagnetic exchange coupling yields the phase diagram with diverse ordering tendencies, while for large value of the coupling the phase diagram has relatively simple constitution.

In the absence of the exchange coupling $J$ and for small nearest-neighbor repulsion $V$, the competition between the chiral $d$-wave SC and the SDW becomes a main ingredient of the phase diagram. The former emerges slightly away from the VHS, while the latter right around it. More specifically, the four-sublattice SDW phase occurs very near the VHS filling, flanked by the incommensurate SDW. When increasing $V$, the spin-triplet $f$-wave SC and the incommensurate CDW phases occur at lower doping levels. If $V$ is further increased, the CDW phase, with broken $\pi /3$-rotation symmetry and a charge transfer between two sublattices, is preferred for all doping levels considered. The associated critical scales are considerably increased. When a weak exchange coupling is included, the structure of the phase diagram changes a lot. The chiral $d$-wave SC, found in the absence of $J$, completely disappears from our phase diagram and the region of the $f$-wave SC is extended. If the exchange coupling is further increased, the four-sublattice SDW turns into the ferromagnetic or incommensurate SDW phases. When the ferromagnetic exchange interaction is further enhanced, the phase diagrams have simple structures consisting only of two phases, namely, the CDW and FM phases. With increasing $J$, the region of the ferromagnetic phase gets extended more and more, ultimately leading to the whole phase diagram covered by the phase.

From the experimental point of view, some previous works have reported the experimental observations of SC in single layer graphene by doping it with Li adatoms \cite{ref49} or intercalating graphene laminates with Ca \cite{ref50}. However, those are different from the unconventional SC addressed in this work, because the doping levels are much lower than the VHS filling and they are the conventional BCS superconductors mediated by dopant-enhanced electron-phonon coupling. Another works on graphene have also found the conventional SC obeying the BCS mechanism \cite{ref51} and a $p$-wave unconventional SC \cite{ref52}, triggered by placing graphene on a superconductor. These proximity-induced SCs are far away from the present context. Thus, the unconventional SC, which was predicted theoretically more than a decade ago for graphene doped close to the VHS, has not yet been found experimentally. Our theoretical finding, that demonstrates a strong suppression of the chiral $d$-wave SC by weak exchange coupling of $J \approx 0.14{\rm{eV}}$, might help to present a key to explain the reason for a failure of the experimental effort for finding the chiral SC in single layer graphene.

Lastly, we give a brief comment on the relevance of our results to the TBG. As mentioned in the introduction of present paper, a number of studies on TBG \cite{ref04, ref05, ref06, ref07, ref08, ref09, ref10, ref57, ref58, ref59, ref60} have pointed to the chiral $d$-wave SC as the nature of the SC state observed in the system. Due to the large lattice constant in TBG, it is expected that the ratio between the radius of the maximally localized Wannier orbitals ($\alpha$) and the nearest-neighbor distance ($R$) for the system would be much smaller than that for single layer graphene. On the other hand, the ratio between the exchange integral and the hopping parameter, $J/t$ decays exponentially with $\xi  = \alpha /R$, as shown in Fig. \ref{fig11} of Appendix \ref{appendA}. Thus, the effect of the ferromagnetic exchange on the many-body instabilities can be neglected, and the theoretically predicted chiral $d$-wave SC may survive in TBG, unlike in single layer graphene.\\

\section*{acknowledgments}
We thank Chol-Jun Kang and Kwang-Il Ryom for useful discussions.

\appendix

\begin{widetext}
	
\section{Estimation of exchange interaction} \label{appendA}

We present a rough estimate of the strength of the exchange interaction for graphene. The Coulomb interaction Hamiltonian can be expanded in terms of the Wannier orbitals.
\begin{eqnarray} \label{eqA1}
\begin{split}
H_{{\mathop{\rm int}} }  =& \frac{1}{2}\int {d{\bf{r}}_1 \int {d{\bf{r}}_2 \sum\limits_{\sigma ,\sigma '} {\psi _\sigma ^ \dag  ({\bf{r}}_1 )\psi _{\sigma '}^ \dag ({\bf{r}}_2 )\psi _{\sigma '} ({\bf{r}}_2 )\psi _\sigma  ({\bf{r}}_1 )} } } \frac{{e^2 }}{{\left| {{\bf{r}}_1  - {\bf{r}}_2 } \right|}}\\
=& \frac{1}{2}\sum\limits_{i'j',ij} {\sum\limits_{o'p',op} {\sum\limits_{\sigma ,\sigma '} {c_{i'o'\sigma }^ \dag  c_{j'p'\sigma '}^ \dag  c_{jp\sigma '} c_{io\sigma } } } } (i'o',j'p'|io,jp),\\
 (i'o',j'p'|io,jp) \equiv & \int {d{\bf{r}}\int {d{\bf{r'}}} } \varphi _{i'o'}^* ({\bf{r}})\varphi _{j'p'}^* ({\bf{r'}})\varphi _{jp} ({\bf{r'}})\varphi _{io} ({\bf{r}})\frac{{e^2 }}{{\left| {{\bf{r}} - {\bf{r'}}} \right|}}.
\end{split}
\end{eqnarray}
Here $\varphi _{io} ({\bf{r}})$ is the maximally localized Wannier function centered at the atom of the sublattice $o$ in the unit cell $i$. Among the integrals $(i'o',j'p'|io,jp)$, the following terms have relatively large values:
\begin{eqnarray}
\nonumber
\begin{split}
U &= (io,io|io,io) = \int {d{\bf{r}}\int {d{\bf{r'}}} } \left| {\varphi _{io} ({\bf{r}})} \right|^2 \frac{{e^2 }}{{\left| {{\bf{r}} - {\bf{r'}}} \right|}}\left| {\varphi _{io} ({\bf{r'}})} \right|^2 \hspace{1.3pc} (\text{on-site repulsion}),\\
V_{io,jp} &= (io,jp|io,jp) = \int {d{\bf{r}}\int {d{\bf{r'}}} } \left| {\varphi _{io} ({\bf{r}})} \right|^2 \frac{{e^2 }}{{\left| {{\bf{r}} - {\bf{r'}}} \right|}}\left| {\varphi _{jp} ({\bf{r'}})} \right|^2 \hspace{1.3pc} (\text{repulsion between neighbors}),\\
J_{io,jp} &= (io,jp|jp,io) = \int {d{\bf{r}}\int {d{\bf{r'}}} } [\varphi _{jp}^* ({\bf{r}})\varphi _{io} ({\bf{r}})]^* \frac{{e^2 }}{{\left| {{\bf{r}} - {\bf{r'}}} \right|}}[\varphi _{jp}^* ({\bf{r'}})\varphi _{io} ({\bf{r'}})] \hspace{1.3pc} (\text{direct exchange interaction}),\\
K_{io,jp} &= (io,io|jp,jp) = \int {d{\bf{r}}\int {d{\bf{r'}}} } [\varphi _{io}^* ({\bf{r}})\varphi _{jp} ({\bf{r}})]\frac{{e^2 }}{{\left| {{\bf{r}} - {\bf{r'}}} \right|}}[\varphi _{io}^* ({\bf{r'}})\varphi _{jp} ({\bf{r'}})] \hspace{1.3pc} (\text{pair hopping}).
\end{split}
\end{eqnarray}
One can take real functions as the Wannier orbitals for graphene's $\pi$-electron system, which gives the relation of $J_{io,jp}  = K_{io,jp}$. In general, many of the above terms are neglected, except for the following parameters:
\begin{eqnarray}
\nonumber
\begin{split}
& \text{on-site repulsion } U, \text{nearest-neighbor Coulomb repulsion } V_{\left\langle {iA,jB} \right\rangle }  = V_{\left\langle {jB,iA} \right\rangle }  = V,\\
& \text{nearest-neighbor exchange coupling } J_{\left\langle {iA,jB} \right\rangle }  = J_{\left\langle {jB,iA} \right\rangle }  = J,\\
& \text{nearest-neighbor pair hopping } K_{\left\langle {iA,jB} \right\rangle }  = K_{\left\langle {jB,iA} \right\rangle }  = J.
\end{split}
\end{eqnarray}
\end{widetext}
In this case the Hamiltonian $H_{{\mathop{\rm int}} }$ is approximated as
\begin{eqnarray} \label{eqA2}
\begin{split}
H_{{\rm{int}}}  \approx {\kern 3pt} & U \sum\limits_{i,o} {n_{io \uparrow } } n_{io \downarrow }  + V\sum\limits_{\left\langle {iA,jB} \right\rangle } {\sum\limits_{\sigma ,\sigma '} {n_{iA\sigma } } } n_{jB\sigma '} \hspace{2.5pc} \\
& + J\sum\limits_{\left\langle {iA,jB} \right\rangle } {\sum\limits_{\sigma ,\sigma '} {c_{iA\sigma }^ \dag  c_{jB\sigma '}^ \dag  } } c_{iA\sigma '} c_{jB\sigma } \\
& + J\sum\limits_{\left\langle {iA,jB} \right\rangle } {(c_{iA \uparrow }^ \dag  c_{iA \downarrow }^ \dag  } c_{jB \downarrow } c_{jB \uparrow }  + {\rm{H}}{\rm{.c}}{\rm{.}}),
\end{split}
\end{eqnarray}
where $n_{io\sigma }  = c_{io\sigma }^ \dag  c_{io\sigma }$ is the local electron density operator for spin polarity $\sigma$.

It is easy to prove that the nearest-neighbor exchange coupling $J$ is always positive. The coupling is expressed using a real function $F({\bf{r}}) = \varphi _{jB} ({\bf{r}})\varphi _{iA} ({\bf{r}})$ as
\begin{eqnarray} \label{eqA3}
\begin{split}
J = J_{\left\langle {iA,jB} \right\rangle }  = \int {d{\bf{r}}\int {d{\bf{r'}}} } F({\bf{r}})\frac{{e^2 }}{{\left| {{\bf{r}} - {\bf{r'}}} \right|}}F({\bf{r'}}). \hspace{1.5pc} 
\end{split}
\end{eqnarray}
The Fourier transforms of $F({\bf{r}})$ and $\frac{{e^2 }}{{\left| {{\bf{r}} - {\bf{r'}}} \right|}}$ are given by
\begin{eqnarray}
\nonumber
\begin{split}
F({\bf{r}}) &= \frac{1}{{(2\pi )^3 }}\int {d{\bf{k}}} \tilde F({\bf{k}})e^{i{\bf{k}} \cdot {\bf{r}}},\\
 \frac{{e^2 }}{{\left| {{\bf{r}} - {\bf{r'}}} \right|}} &= \frac{1}{{(2\pi )^3 }}\int {d{\bf{k}}} \frac{{4\pi e^2 }}{{k^2 }}e^{i{\bf{k}} \cdot ({\bf{r}} - {\bf{r'}})}.
\end{split}
\end{eqnarray}
By substituting these into Eq. (\ref{eqA3}) we obtain
\begin{eqnarray} \label{eqA4}
\begin{split}
J =& \frac{1}{{(2\pi )^3 }}\int {d{\bf{k}}} \tilde F( - {\bf{k}})\frac{{4\pi e^2 }}{{k^2 }}\tilde F({\bf{k}}) \\
=& \frac{1}{{(2\pi )^3 }}\int {d{\bf{k}}} \frac{{4\pi e^2 }}{{k^2 }}\left| {\tilde F({\bf{k}})} \right|^2  > 0.
\end{split}
\end{eqnarray}
Here we used the relation for any real function $F({\bf{r}})$, given by $\tilde F(-{\bf{k}}) = [\tilde F({\bf{k}})]^*$. Thus, the exchange integral $J$ has always positive value.

The third term in Eq. (\ref{eqA2}) can be represented in the following form:
\begin{eqnarray} \label{eqA5}
\begin{split}
H_{{\rm{exc}}} & \equiv J\sum\limits_{\left\langle {iA,jB} \right\rangle } {\sum\limits_{\sigma ,\sigma '} {c_{iA\sigma }^ \dag  c_{jB\sigma '}^ \dag  c_{iA\sigma '} c_{jB\sigma } } } \\
& =  - \frac{J}{2}\sum\limits_{\left\langle {iA,jB} \right\rangle } {n_{iA} n_{jB} }  - 2J\sum\limits_{\left\langle {iA,jB} \right\rangle } {{\bf{S}}_{iA}  \cdot {\bf{S}}_{jB} } . \hspace{1.5pc} 
\end{split}
\end{eqnarray}
The first term can be added to the nearest-neighbor Coulomb repulsion, while the second one is just the ferromagnetic ($J>0$) exchange interaction with the tendency to align the spin orientations.

Now we consider graphene's $\pi$ electrons that move in the effective field produced by ion cores and $\sigma$ electrons. As a crude approximation we assume that this effective field is identical with the one produced by an array (the honeycomb lattice) of effective point charges of strength $+Qe$. The parameter $Q$ has a value in the range of 1 (complete screening by $\sigma$ electrons) to 4 (no screening). The $2p_z$ orbital of an electron moving in the Coulomb attraction of a point charge $Qe$ at the origin reads
\begin{eqnarray} \label{eqA6}
\begin{split}
f({\bf{r}}) = \frac{1}{{\sqrt {\pi \alpha ^3 } }}\frac{z}{\alpha }e^{ - r/\alpha }, \alpha  = \frac{{2a_B }}{Q},
\end{split}
\end{eqnarray}
where $a_B = 0.53 \rm{\AA}$ is the Bohr radius.

On the other hand, in the case of graphene's $\pi$ electrons, the maximally localized Wannier orbital centered at the site $i$, which should be orthogonal to the others, can be approximately written as
\begin{eqnarray} \label{eqA7}
\begin{split}
\varphi _i ({\bf{r}}) \approx \frac{1}{{\sqrt {1 - \frac{9}{4}S^2 } }}\left[ {f({\bf{r}} - {\bf{R}}_i ) - \frac{S}{2}\sum\limits_{j = 1}^3 {f({\bf{r}} - {\bf{R}}_j )} } \right]. \hspace{1.8pc}
\end{split}
\end{eqnarray}
Here $S = \left\langle {{f_A }} \mathrel{|} {{f_B }} \right\rangle  = \int {d{\bf{r}}} f({\bf{r}} - {\bf{R}}_A )f({\bf{r}} - {\bf{R}}_B )$ is the overlap between the nearest-neighboring atomic orbitals, ${\bf{R}}_i$ is the position vector of the site $i$, and the sites $j = 1,2,3$ are three nearest neighbors of the site $i$. In deriving Eq. (\ref{eqA7}), we assumed the smallness of the overlap $S$ and took into account the equivalence of all atoms. By using Eqs. (\ref{eqA6}) and (\ref{eqA7}) one can obtain the exchange coupling $J$:
\begin{widetext}
\begin{eqnarray} \label{eqA8}
\begin{split}
J &= \int {d{\bf{r}}\int {d{\bf{r'}}} } \varphi _A ({\bf{r}})\varphi _B ({\bf{r}})\frac{{e^2 }}{{\left| {{\bf{r}} - {\bf{r'}}} \right|}}\varphi _A ({\bf{r'}})\varphi _B ({\bf{r'}})\\
&\approx \frac{1}{{\left( {1 - \frac{9}{4}S^2 } \right)^2 }}\left[ {\left( {1 + \frac{{S^2 }}{4}} \right)^2 J_{AB}  + \frac{{S^2 }}{2}(U + V_{AB} ) - 2S\left( {1 + \frac{{S^2 }}{4}} \right)X_{AB - A} } \right]\\
&\approx \left( {1 + \frac{9}{4}S^2 } \right)^2 \left[ {\left( {1 + \frac{{S^2 }}{4}} \right)^2 J_{AB}  + \frac{{S^2 }}{2}(U + V_{AB} ) - 2S\left( {1 + \frac{{S^2 }}{4}} \right)X_{AB - A} } \right].
\end{split}
\end{eqnarray}
In the above equation, the integrals $J_{AB} ,U,V_{AB}$, and $X_{AB - A}$ are defined by atomic orbitals $f_A  = f({\bf{r}} - {\bf{R}}_A )$ and $f_B  = f({\bf{r}} - {\bf{R}}_B )$ as follows:
\begin{eqnarray} \label{eqA9}
\begin{split}
J_{AB}  &\equiv \int {d{\bf{r}}\int {d{\bf{r'}}} } f_A ({\bf{r}})f_B ({\bf{r}})\frac{{e^2 }}{{\left| {{\bf{r}} - {\bf{r'}}} \right|}}f_A ({\bf{r'}})f_B ({\bf{r'}}), U \equiv \int {d{\bf{r}}\int {d{\bf{r'}}} } f^2 ({\bf{r}})\frac{{e^2 }}{{\left| {{\bf{r}} - {\bf{r'}}} \right|}}f^2 ({\bf{r'}}),\\
V_{AB}  &\equiv \int {d{\bf{r}}\int {d{\bf{r'}}} } f_A^2 ({\bf{r}})\frac{{e^2 }}{{\left| {{\bf{r}} - {\bf{r'}}} \right|}}f_B^2 ({\bf{r'}}), X_{AB - A}  \equiv \int {d{\bf{r}}\int {d{\bf{r'}}} } f_A ({\bf{r}})f_B ({\bf{r}})\frac{{e^2 }}{{\left| {{\bf{r}} - {\bf{r'}}} \right|}}f_A^2 ({\bf{r'}}).
\end{split}
\end{eqnarray}
For the distance between nearest neighbors of $R$, the overlap $S$ is given by
\begin{eqnarray} \label{eqA10}
\begin{split}
S = e^{ - R/\alpha } \left( {1 + \frac{R}{\alpha } + \frac{2}{5}\frac{{R^2 }}{{\alpha ^2 }} + \frac{1}{{15}}\frac{{R^3 }}{{\alpha ^3 }}} \right).
\end{split}
\end{eqnarray}
The hopping amplitude between two nearest-neighboring Wannier orbitals is written, to first order in $S$, as
\begin{eqnarray} \label{eqA11}
\begin{split}
t &=  - \left\langle {\varphi _A } \right|\hat H\left| {\varphi _B } \right\rangle  \approx  - \left\langle {f_A } \right|\hat H\left| {f_B } \right\rangle  + S\left\langle {f_A } \right|\hat H\left| {f_A } \right\rangle\\
&\approx  - \frac{1}{2}\int {d{\bf{r}}} f({\bf{r}} - {\bf{R}}_A )[V_A ({\bf{r}}) + V_B ({\bf{r}})]f({\bf{r}} - {\bf{R}}_B ) + S\int {d{\bf{r}}} V_B ({\bf{r}})f^2 ({\bf{r}} - {\bf{R}}_A )\\
&\approx Q\frac{{e^2 }}{{2\alpha }}e^{ - R/\alpha } \left( {1 + \frac{R}{\alpha } + \frac{1}{3}\frac{{R^2 }}{{\alpha ^2 }}} \right) - SQ\frac{{e^2 }}{R}\left( {1 - \frac{3}{2}\frac{{\alpha ^2 }}{{R^2 }}} \right),
\end{split}
\end{eqnarray}
\end{widetext}
with the potential by the effective point charge $Qe$ at the position ${\bf{R}}_i$, $V_i ({\bf{r}}) \equiv  - Q\frac{{e^2 }}{{\left| {{\bf{r}} - {\bf{R}}_i } \right|}}$.

The on-site repulsion $U$ for an atomic orbital in Eq. (\ref{eqA9}) is approximately equal to that for a Wannier orbital. So we can extract the parameters $\alpha$ and $Q$ from $U$, from which the exchange coupling $J$ and the hopping parameter $t$ are calculated using Eqs. (\ref{eqA8}) and (\ref{eqA11}). In Ref. \onlinecite{ref53} two values of on-site repulsions, $U_{{\rm{bare}}}  = 17.0{\rm{eV}}$ and $U_{{\rm{cRPA}}}  = 9.3{\rm{eV}}$, have been obtained using both the bare Coulomb interaction and the screened interaction from the constrained random phase approximation (cRPA), respectively, while in this paper we used the parameter, $U_{{\rm{used}}}  = 3.6t = 3.6 \times 2.8{\rm{eV}} \approx 10.1{\rm{eV}}$. Since graphene has the C-C distance of $R = 1.42 \rm{\AA{}}$, we get the results for these three values of $U$ as shown in Table \ref{tab02}.

\begin{table}[h]
	\caption{The calculation results of the hopping parameters and the exchange couplings for three values of on-site repulsions.}
	\begin{center}
		\begin{tabular}{|c|c|c|c|c|c|c|}
			\hline
			$U$ (eV) & $\alpha$ (\AA{}) & $\xi  = R/\alpha$ & $Q$ & $S$  & $t$ (eV) & $J$ (eV)\\		 
			\hline
			9.3 & 0.61 & 2.33 & 1.74 & 0.62  & 2.39 & 0.28 \\		 
			10.1 & 0.56 & 2.54 & 1.89 & 0.57 & 2.56 & 0.25 \\		 
			17.0 & 0.33 & 4.30 & 3.21 & 0.24 & 3.58 & 0.10 \\		 
			\hline
		\end{tabular}
	\end{center}
	\label{tab02}
\end{table}

\begin{figure}[h]
	\begin{center}
		\includegraphics[width=8.0cm]{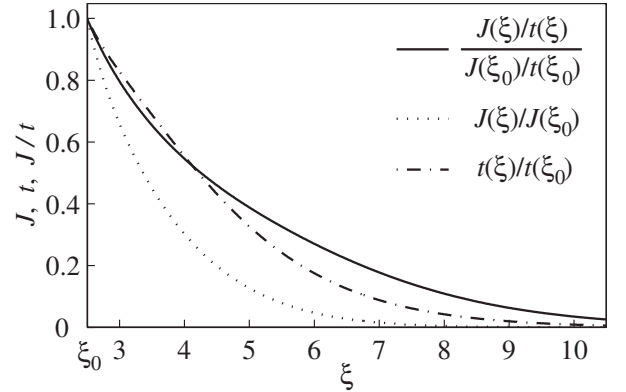}
	\end{center}
	\caption{The dependence of the exchange coupling $J$ and the hopping amplitude $t$ on the parameter $\xi  = R/\alpha$ ($\xi _0  = 2.5$).}
	\label{fig11}
\end{figure}

As a consequence, we anticipate the strength of exchange coupling to be $0.10 \sim 0.28 \rm{eV}$, from the estimation using three legitimate values of the parameter $U$. Fig. \ref{fig11} demonstrates the dependence of $J$, $t$, and $J/t$ on the parameter $\xi  = R/\alpha$. One can easily see from the figure that the quantity $J/t$ decays exponentially with $\xi$, implying a negligible effect of the exchange interaction in the case of the superlattices with large lattice constants.

\begin{widetext}

\section{Projections onto three channels} \label{appendB}

Three bosonic propagators are defined by projecting three single-channel coupling functions onto three associated channels [see Eq. (\ref{eq09})]:
\begin{eqnarray} \label{eqB1}
\begin{split}
P^\Omega   = {\rm{\hat P}}[\Phi ^{{\rm{pp}}} (\Omega )],C^\Omega   = {\rm{\hat C}}[\Phi ^{{\rm{ph,cr}}} (\Omega )],D^\Omega   = {\rm{\hat D}}[\Phi ^{{\rm{ph,d}}} (\Omega )].
\end{split}
\end{eqnarray}
Their detailed expressions are given by
\begin{eqnarray} \label{eqB2}
\begin{split}
P_{o'_1 o'_2 m,o_1 o_2 n}^\Omega  ({\bf{q}}) &= \frac{1}{{S_{BZ}^2 }}\int {d{\bf{p}}} \int {d{\bf{p}}'} f_m ({\bf{p}})f_n^* ({\bf{p}}')\Phi _{o'_1 o'_2 ,o_1 o_2 }^{{\rm{pp}}(\Omega )} ({\bf{p}} + {\bf{q}}, - {\bf{p}};{\bf{p}}' + {\bf{q}}, - {\bf{p}}'),\\
C_{o'_1 o_2 m,o_1 o'_2 n}^\Omega  ({\bf{q}}) &= \frac{1}{{S_{BZ}^2 }}\int {d{\bf{p}}} \int {d{\bf{p}}'} f_m ({\bf{p}})f_n^* ({\bf{p}}')\Phi _{o'_1 o'_2 ,o_1 o_2 }^{{\rm{ph,cr}}(\Omega )} ({\bf{p}} + {\bf{q}},{\bf{p}}';{\bf{p}}' + {\bf{q}},{\bf{p}}),\\
D_{o'_1 o_1 m,o_2 o'_2 n}^\Omega  ({\bf{q}}) &= \frac{1}{{S_{BZ}^2 }}\int {d{\bf{p}}} \int {d{\bf{p}}'} f_m ({\bf{p}})f_n^* ({\bf{p}}')\Phi _{o'_1 o'_2 ,o_1 o_2 }^{{\rm{ph,d}}(\Omega )} ({\bf{p}} + {\bf{q}},{\bf{p}}';{\bf{p}},{\bf{p}}' + {\bf{q}}).
\end{split}
\end{eqnarray}
The inverse transformations of above equation read as follows:
\begin{eqnarray} \label{eqB3}
\begin{split}
\Phi _{o'_1 o'_2 ,o_1 o_2 }^{{\rm{pp(}}\Omega {\rm{)}}} ({\bf{p}} + {\bf{q}}, - {\bf{p}};{\bf{k}} + {\bf{q}}, - {\bf{k}}) &= \sum\limits_{m,n{\rm{(infinit}}{\kern 1pt} {\kern 1pt} {\kern 1pt} {\rm{sum}})} {P_{o'_1 o'_2 m,o_1 o_2 n}^\Omega  } ({\bf{q}})f_m^* ({\bf{p}})f_n ({\bf{k}}),\\
\Phi _{o'_1 o'_2 ,o_1 o_2 }^{{\rm{ph,cr(}}\Omega {\rm{)}}} ({\bf{p}} + {\bf{q}},{\bf{k}};{\bf{k}} + {\bf{q}},{\bf{p}}) &= \sum\limits_{m,n{\rm{(infinit}}{\kern 1pt} {\kern 1pt} {\kern 1pt} {\rm{sum}})} {C_{o'_1 o_2 m,o_1 o'_2 n}^\Omega  } ({\bf{q}})f_m^* ({\bf{p}})f_n ({\bf{k}}),\\
\Phi _{o'_1 o'_2 ,o_1 o_2 }^{{\rm{ph,d}}(\Omega )} ({\bf{p}} + {\bf{q}},{\bf{k}};{\bf{p}},{\bf{k}} + {\bf{q}}) &= \sum\limits_{m,n{\rm{(infinit}}{\kern 1pt} {\kern 1pt} {\kern 1pt} {\rm{sum}})} {D_{o'_1 o_1 m,o_2 o'_2 n}^\Omega  } ({\bf{q}})f_m^* ({\bf{p}})f_n ({\bf{k}}),
\end{split}
\end{eqnarray}
which can be shortly represented as
\begin{eqnarray} \label{eqB4}
\begin{split}
\Phi ^{{\rm{pp}}} (\Omega ) = {\rm{\hat P}}^{ - 1} [P^\Omega  ],\Phi ^{{\rm{ph,cr}}} (\Omega ) = {\rm{\hat C}}^{ - 1} [C^\Omega  ],\Phi ^{{\rm{ph,d}}} (\Omega ) = {\rm{\hat D}}^{ - 1} [D^\Omega].
\end{split}
\end{eqnarray}
If the range of the indices $m$ and $n$ (i.e., ${\bf{R}}_m$ and ${\bf{R}}_n$) extend to infinity, Eqs. (\ref{eqB3}) and (\ref{eqB4}) would be exact. However, the real calculation will necessarily introduce the truncation in the range of ${\bf{R}}_m$ and ${\bf{R}}_n$ [see Fig. \ref{fig1}(a)], thus making these equations to be approximate. More specifically, since the bosonic propagators for $\left| {{\bf{R}}_m } \right| > R_{{\rm{cut}}}$ or $\left| {{\bf{R}}_n } \right| > R_{{\rm{cut}}}$ are neglected,
\begin{equation}
\nonumber
P_{o'_1 o'_2 m,o_1 o_2 n}^\Omega  ({\bf{q}}) = C_{o'_1 o'_2 m,o_1 o_2 n}^\Omega  ({\bf{q}}) = D_{o'_1 o'_2 m,o_1 o_2 n}^\Omega  ({\bf{q}}) = 0 \hspace{2pc}  (\left| {{\bf{R}}_m } \right| > R_{{\rm{cut}}} \text{ or } \left| {{\bf{R}}_n } \right| > R_{{\rm{cut}}}),
\end{equation}
Eqs. (\ref{eqB3}) and (\ref{eqB4}) become the following approximations:
\begin{eqnarray} \label{eqB5}
\begin{split}
\Phi _{o'_1 o'_2 ,o_1 o_2 }^{{\rm{pp(}}\Omega {\rm{)}}} ({\bf{p}} + {\bf{q}}, - {\bf{p}};{\bf{k}} + {\bf{q}}, - {\bf{k}})
&\approx \sum\limits_{m,n{\rm{(truncated}}{\kern 1pt} {\kern 1pt} {\kern 1pt} {\rm{sum}})} {P_{o'_1 o'_2 m,o_1 o_2 n}^\Omega  ({\bf{q}})} f_m^* ({\bf{p}})f_n ({\bf{k}}),\\
\Phi _{o'_1 o'_2 ,o_1 o_2 }^{{\rm{ph,cr(}}\Omega {\rm{)}}} ({\bf{p}} + {\bf{q}},{\bf{k}};{\bf{k}} + {\bf{q}},{\bf{p}}) 
&\approx \sum\limits_{m,n{\rm{(truncated}}{\kern 1pt} {\kern 1pt} {\kern 1pt} {\rm{sum}})} {C_{o'_1 o_2 m,o_1 o'_2 n}^\Omega  ({\bf{q}})} f_m^* ({\bf{p}})f_n ({\bf{k}}),\\
\Phi _{o'_1 o'_2 ,o_1 o_2 }^{{\rm{ph,d}}(\Omega )} ({\bf{p}} + {\bf{q}},{\bf{k}};{\bf{p}},{\bf{k}} + {\bf{q}}) 
&\approx \sum\limits_{m,n{\rm{(truncated}}{\kern 1pt} {\kern 1pt} {\kern 1pt} {\rm{sum}})} {D_{o'_1 o_1 m,o_2 o'_2 n}^\Omega  ({\bf{q}})} f_m^* ({\bf{p}})f_n ({\bf{k}}).
\end{split}
\end{eqnarray}
\begin{eqnarray} \label{eqB6}
\begin{split}
\Phi ^{{\rm{pp}}} (\Omega ) \approx {\rm{\hat P}}^{ - 1} [P^\Omega  ],\Phi ^{{\rm{ph,cr}}} (\Omega ) \approx {\rm{\hat C}}^{ - 1} [C^\Omega  ],\Phi ^{{\rm{ph,d}}} (\Omega ) \approx {\rm{\hat D}}^{ - 1} [D^\Omega].
\end{split}
\end{eqnarray}
In Eq. (\ref{eqB5}) the sum $\sum\limits_{m,n{\rm{(truncated}}{\kern 1pt} {\kern 1pt} {\kern 1pt} {\rm{sum}})}$ means $\sum\limits_{m\left( {\left| {{\bf{R}}_m } \right| \le R_{{\rm{cut}}} } \right){\kern 1pt} {\kern 1pt} {\kern 1pt} {\kern 1pt} {\kern 1pt} } {\sum\limits_{n\left( {\left| {{\bf{R}}_n } \right| \le R_{{\rm{cut}}} } \right)}}$.

\section{Crossed contributions to three projection matrices} \label{appendC}

The projection matrices are defined by [see Eq. (\ref{eq11})]
\begin{eqnarray} \label{eqC1}
\begin{split}
V^{\rm{P}} (\Omega ) = {\rm{\hat P}}[V^\Omega  ],V^{\rm{C}} (\Omega ) = {\rm{\hat C}}[V^\Omega  ],V^{\rm{D}} (\Omega ) = {\rm{\hat D}}[V^\Omega],
\end{split}
\end{eqnarray}
which can be represented in detail as
\begin{eqnarray} \label{eqC2}
\begin{split}
V_{o'_1 o'_2 m,o_1 o_2 n}^{{\rm{P}}(\Omega )} ({\bf{q}}) &= \frac{1}{{S_{BZ}^2 }}\int {d{\bf{p}}} \int {d{\bf{p'}}} f_m ({\bf{p}})f_n^* ({\bf{p'}})V_{o'_1 o'_2 ,o_1 o_2 }^\Omega  ({\bf{p}} + {\bf{q}}, - {\bf{p}};{\bf{p'}} + {\bf{q}}, - {\bf{p'}}),\\
V_{o'_1 o_2 m,o_1 o'_2 n}^{{\rm{C}}(\Omega )} ({\bf{q}}) &= \frac{1}{{S_{BZ}^2 }}\int {d{\bf{p}}} \int {d{\bf{p'}}} f_m ({\bf{p}})f_n^* ({\bf{p'}})V_{o'_1 o'_2 ,o_1 o_2 }^\Omega  ({\bf{p}} + {\bf{q}},{\bf{p'}};{\bf{p'}} + {\bf{q}},{\bf{p}}),\\
V_{o'_1 o_1 m,o_2 o'_2 n}^{{\rm{D}}(\Omega )} ({\bf{q}}) &= \frac{1}{{S_{BZ}^2 }}\int {d{\bf{p}}} \int {d{\bf{p'}}} f_m ({\bf{p}})f_n^* ({\bf{p'}})V_{o'_1 o'_2 ,o_1 o_2 }^\Omega  ({\bf{p}} + {\bf{q}},{\bf{p'}};{\bf{p}},{\bf{p'}} + {\bf{q}}).
\end{split}
\end{eqnarray}
The inverse transformations of above equation read as follows:
\begin{eqnarray} \label{eqC3}
\begin{split}
V_{o'_1 o'_2 ,o_1 o_2 }^\Omega  ({\bf{p}} + {\bf{q}}, - {\bf{p}};{\bf{k}} + {\bf{q}}, - {\bf{k}}) &\approx \sum\limits_{m,n{\rm{(truncated}}{\kern 1pt} {\kern 1pt} {\kern 1pt} {\rm{sum}})} {V_{o'_1 o'_2 m,o_1 o_2 n}^{{\rm{P}}(\Omega )} } ({\bf{q}})f_m^* ({\bf{p}})f_n ({\bf{k}}),\\
V_{o'_1 o'_2 ,o_1 o_2 }^\Omega  ({\bf{p}} + {\bf{q}},{\bf{k}};{\bf{k}} + {\bf{q}},{\bf{p}}) &\approx \sum\limits_{m,n{\rm{(truncated}}{\kern 1pt} {\kern 1pt} {\kern 1pt} {\rm{sum}})} {V_{o'_1 o_2 m,o_1 o'_2 n}^{{\rm{C}}(\Omega )} } ({\bf{q}})f_m^* ({\bf{p}})f_n ({\bf{k}}),\\
V_{o'_1 o'_2 ,o_1 o_2 }^\Omega  ({\bf{p}} + {\bf{q}},{\bf{k}};{\bf{p}},{\bf{k}} + {\bf{q}}) &\approx \sum\limits_{m,n{\rm{(truncated}}{\kern 1pt} {\kern 1pt} {\kern 1pt} {\rm{sum}})} {V_{o'_1 o_1 m,o_2 o'_2 n}^{{\rm{D}}(\Omega )} } ({\bf{q}})f_m^* ({\bf{p}})f_n ({\bf{k}}),
\end{split}
\end{eqnarray}
which can be briefly represented as
\begin{eqnarray} \label{eqC4}
\begin{split}
V^\Omega   \approx {\rm{\hat P}}^{ - 1} [V^{\rm{P}} (\Omega )] \approx {\rm{\hat C}}^{ - 1} [V^{\rm{C}} (\Omega )] \approx {\rm{\hat D}}^{ - 1} [V^{\rm{D}} (\Omega )].
\end{split}
\end{eqnarray}

On the other hand, the effective interaction is represented via the bosonic propagators as [see Eq. (\ref{eq25})]
\begin{eqnarray} \label{eqC5}
\begin{split}
V^\Omega   = V^{(0)}  + \Phi ^{{\rm{pp}}} (\Omega ) + \Phi ^{{\rm{ph,cr}}} (\Omega ) + \Phi ^{{\rm{ph,d}}} (\Omega )
\approx V^{(0)}  + {\rm{\hat P}}^{ - 1} [P^\Omega  ] + {\rm{\hat C}}^{ - 1} [C^\Omega  ] + {\rm{\hat D}}^{ - 1} [D^\Omega],
\end{split}
\end{eqnarray}
from which the projection matrices are obtained:
\begin{eqnarray} \label{eqC6}
\begin{split}
V^{\rm{P}} (\Omega ) &= {\rm{\hat P}}[V^{(0)} ] + {\rm{\hat P}}[\Phi ^{{\rm{pp}}} (\Omega )] + {\rm{\hat P}}[\Phi ^{{\rm{ph,cr}}} (\Omega )] + {\rm{\hat P}}[\Phi ^{{\rm{ph,d}}} (\Omega )]
= V^{{\rm{P}},(0)}  + P^\Omega   + V^{{\rm{P}} \leftarrow {\rm{C}}} (\Omega ) + V^{{\rm{P}} \leftarrow {\rm{D}}} (\Omega ),\\
V^{\rm{C}} (\Omega ) &= {\rm{\hat C}}[V^{(0)} ] + {\rm{\hat C}}[\Phi ^{{\rm{pp}}} (\Omega )] + {\rm{\hat C}}[\Phi ^{{\rm{ph,cr}}} (\Omega )] + {\rm{\hat C}}[\Phi ^{{\rm{ph,d}}} (\Omega )]
= V^{{\rm{C}},(0)}  + V^{{\rm{C}} \leftarrow {\rm{P}}} (\Omega ) + C^\Omega   + V^{{\rm{C}} \leftarrow {\rm{D}}} (\Omega ),\\
V^{\rm{D}} (\Omega ) &= {\rm{\hat D}}[V^{(0)} ] + {\rm{\hat D}}[\Phi ^{{\rm{pp}}} (\Omega )] + {\rm{\hat D}}[\Phi ^{{\rm{ph,cr}}} (\Omega )] + {\rm{\hat D}}[\Phi ^{{\rm{ph,d}}} (\Omega )]
= V^{{\rm{D}},(0)}  + V^{{\rm{D}} \leftarrow {\rm{P}}} (\Omega ) + V^{{\rm{D}} \leftarrow {\rm{C}}} (\Omega ) + D^\Omega  .
\end{split}
\end{eqnarray}
Here $V^{{\rm{P}},(0)} ,V^{{\rm{C}},(0)} ,V^{{\rm{D}},(0)} ,V^{{\rm{P}} \leftarrow {\rm{C}}} ,V^{{\rm{P}} \leftarrow {\rm{D}}} ,V^{{\rm{C}} \leftarrow {\rm{P}}} ,V^{{\rm{C}} \leftarrow {\rm{D}}} ,V^{{\rm{D}} \leftarrow {\rm{P}}}$, and $V^{{\rm{D}} \leftarrow {\rm{C}}}$ are defined by
\begin{eqnarray} \label{eqC7}
\begin{split}
&V^{{\rm{P}},(0)}  \equiv {\rm{\hat P}}[V^{(0)} ],V^{{\rm{C}},(0)}  \equiv {\rm{\hat C}}[V^{(0)} ],V^{{\rm{D}},(0)}  \equiv {\rm{\hat D}}[V^{(0)} ],\\
&V^{{\rm{P}} \leftarrow {\rm{C}}} (\Omega ) \equiv {\rm{\hat P}}[\Phi ^{{\rm{ph,cr}}} (\Omega )] \approx {\rm{\hat P}}\{ {\rm{\hat C}}^{ - 1} [C^\Omega  ]\} ,V^{{\rm{P}} \leftarrow {\rm{D}}} (\Omega ) \equiv {\rm{\hat P}}[\Phi ^{{\rm{ph,d}}} (\Omega )] \approx {\rm{\hat P}}\{ {\rm{\hat D}}^{ - 1} [D^\Omega  ]\} ,\\
&V^{{\rm{C}} \leftarrow {\rm{P}}} (\Omega ) \equiv {\rm{\hat C}}[\Phi ^{{\rm{pp}}} (\Omega )] \approx {\rm{\hat C}}\{ {\rm{\hat P}}^{ - 1} [P^\Omega  ]\} ,V^{{\rm{C}} \leftarrow {\rm{D}}} (\Omega ) \equiv {\rm{\hat C}}[\Phi ^{{\rm{ph,d}}} (\Omega )] \approx {\rm{\hat C}}\{ {\rm{\hat D}}^{ - 1} [D^\Omega  ]\} ,\\
&V^{{\rm{D}} \leftarrow {\rm{P}}} (\Omega ) \equiv {\rm{\hat D}}[\Phi ^{{\rm{pp}}} (\Omega )] \approx {\rm{\hat D}}\{ {\rm{\hat P}}^{ - 1} [P^\Omega  ]\} ,V^{{\rm{D}} \leftarrow {\rm{C}}} (\Omega ) \equiv {\rm{\hat D}}[\Phi ^{{\rm{ph,cr}}} (\Omega )] \approx {\rm{\hat D}}\{ {\rm{\hat C}}^{ - 1} [C^\Omega  ]\}.
\end{split}
\end{eqnarray}

The crossed contributions to the projection matrices can be expressed in terms of $P^\Omega$, $C^\Omega$ and $D^\Omega$. As an example, we can represent the crossed contribution $V^{{\rm{P}} \leftarrow {\rm{C}}} (\Omega)$ via the bosonic propagators. From the relation $V^{{\rm{P}} \leftarrow {\rm{C}}} (\Omega ) = {\rm{\hat P}}[\Phi ^{{\rm{ph,cr}}} (\Omega)]$, we have
\begin{equation}
\nonumber
V_{o'_1 o'_2 m,o_1 o_2 n}^{{\rm{P}} \leftarrow {\rm{C}}(\Omega )} ({\bf{q}}) = \frac{1}{{S_{BZ}^2 }}\int {d{\bf{p}}} \int {d{\bf{p'}}} f_m ({\bf{p}})f_n^* ({\bf{p'}})\Phi _{o'_1 o'_2 ,o_1 o_2 }^{{\rm{ph,cr}}(\Omega )} ({\bf{p}} + {\bf{q}}, - {\bf{p}};{\bf{p'}} + {\bf{q}}, - {\bf{p'}}).
\end{equation}
Substituting Eq. (\ref{eqB5}) into the above equation we obtain the following equation:
\begin{eqnarray}
\nonumber
\begin{split}
V_{o'_1 o'_2 m,o_1 o_2 n}^{{\rm{P}} \leftarrow {\rm{C}}(\Omega )} ({\bf{q}}) &\approx \frac{1}{{S_{BZ}^2 }}\int {d{\bf{p}}} \int {d{\bf{p'}}} f_m ({\bf{p}})f_n^* ({\bf{p'}})\sum\limits_{m',n'} {C_{o'_1 o_2 m',o_1 o'_2 n'}^\Omega  ({\bf{p}} + {\bf{p'}} + {\bf{q}})} f_{m'}^* ( - {\bf{p'}})f_{n'} ( - {\bf{p}})\\
&= \sum\limits_{m',n'} {\frac{1}{{S_{BZ}^2 }}} \int {d{\bf{p}}} \int {d{\bf{p''}}} e^{i{\bf{R}}_m  \cdot {\bf{p}}} e^{ - i{\bf{R}}_n  \cdot ({\bf{p''}} - {\bf{p}} - {\bf{q}})} C_{o'_1 o_2 m',o_1 o'_2 n'}^\Omega  ({\bf{p''}})e^{i{\bf{R}}_{m'}  \cdot ({\bf{p''}} - {\bf{p}} - {\bf{q}})} e^{ - i{\bf{R}}_{n'}  \cdot {\bf{p}}}\\
&= \sum\limits_{l,n'} {e^{i({\bf{R}}_n  - {\bf{R}}_l ) \cdot {\bf{q}}} } \delta _{m + n,l + n'} \frac{1}{{S_{BZ} }}\int {d{\bf{p''}}} C_{o'_1 o_2 l,o_1 o'_2 n'}^\Omega  ({\bf{p''}})e^{ - i({\bf{R}}_n  - {\bf{R}}_l ) \cdot {\bf{p''}}}\\
&= \sum\limits_l {\tilde C_{o'_1 ,o_2 ,{\bf{R}}_l ;o_1 ,o'_2 ,{\bf{R}}_m  + {\bf{R}}_n  - {\bf{R}}_l }^\Omega  ({\bf{R}}_n  - {\bf{R}}_l )} e^{i({\bf{R}}_n  - {\bf{R}}_l ) \cdot {\bf{q}}}.
\end{split}
\end{eqnarray}
The expressions for other crossed contributions can also be derived in a similar way. The results are summarized as follows:
\begin{eqnarray} \label{eqC8}
\begin{split}
V^{\rm{P}} (\Omega ) &= V^{{\rm{P}},(0)}  + P^\Omega   + V^{{\rm{P}} \leftarrow {\rm{C}}} (\Omega ) + V^{{\rm{P}} \leftarrow {\rm{D}}} (\Omega ),\\
V_{o'_1 o'_2 m,o_1 o_2 n}^{{\rm{P}} \leftarrow {\rm{C}}(\Omega )} ({\bf{q}}) &= \sum\limits_l {\tilde C_{o'_1 ,o_2 ,{\bf{R}}_l ;o_1 ,o'_2 ,{\bf{R}}_m  + {\bf{R}}_n  - {\bf{R}}_l }^\Omega  ({\bf{R}}_n  - {\bf{R}}_l )} e^{i({\bf{R}}_n  - {\bf{R}}_l ) \cdot {\bf{q}}} ,\\
V_{o'_1 o'_2 m,o_1 o_2 n}^{{\rm{P}} \leftarrow {\rm{D}}(\Omega )} ({\bf{q}}) &= \sum\limits_l {\tilde D_{o'_1 ,o_1 ,{\bf{R}}_l ;o_2 ,o'_2 ,{\bf{R}}_m  - {\bf{R}}_n  - {\bf{R}}_l }^\Omega  } ( - {\bf{R}}_n  - {\bf{R}}_l )e^{ - i{\bf{R}}_l  \cdot {\bf{q}}},
\end{split}
\end{eqnarray}
\begin{eqnarray} \label{eqC9}
\begin{split}
V^{\rm{C}} (\Omega ) &= V^{{\rm{C}},(0)}  + C^\Omega   + V^{{\rm{C}} \leftarrow {\rm{P}}} (\Omega ) + V^{{\rm{C}} \leftarrow {\rm{D}}} (\Omega ),\\
V_{o'_1 o_2 m,o_1 o'_2 n}^{{\rm{C}} \leftarrow {\rm{P}}(\Omega )} ({\bf{q}}) &= \sum\limits_l {\tilde P_{o'_1 ,o'_2 ,{\bf{R}}_l ;o_1 ,o_2 ,{\bf{R}}_m  + {\bf{R}}_n  - {\bf{R}}_l }^\Omega  ({\bf{R}}_n  - {\bf{R}}_l )e^{i({\bf{R}}_n  - {\bf{R}}_l ) \cdot {\bf{q}}} } ,\\
V_{o'_1 o_2 m,o_1 o'_2 n}^{{\rm{C}} \leftarrow {\rm{D}}(\Omega )} ({\bf{q}}) &= \sum\limits_l {\tilde D_{o'_1 ,o_1 ,{\bf{R}}_l ;o_2 ,o'_2 ,{\bf{R}}_n  + {\bf{R}}_l  - {\bf{R}}_m }^\Omega  ( - {\bf{R}}_m )e^{ - i{\bf{R}}_l  \cdot {\bf{q}}}},
\end{split}
\end{eqnarray}
\begin{eqnarray} \label{eqC10}
\begin{split}
V^{\rm{D}} (\Omega ) &= V^{{\rm{D}},(0)}  + D^\Omega   + V^{{\rm{D}} \leftarrow {\rm{P}}} (\Omega ) + V^{{\rm{D}} \leftarrow {\rm{C}}} (\Omega),\\
V_{o'_1 o_1 m,o_2 o'_2 n}^{{\rm{D}} \leftarrow {\rm{P}}(\Omega )} ({\bf{q}}) &= \sum\limits_l {\tilde P_{o'_1 ,o'_2 ,{\bf{R}}_l ;o_1 ,o_2 ,{\bf{R}}_l  - {\bf{R}}_m  - {\bf{R}}_n }^\Omega  ( - {\bf{R}}_m )e^{i({\bf{R}}_n  - {\bf{R}}_l ) \cdot {\bf{q}}} },\\
V_{o'_1 o_1 m,o_2 o'_2 n}^{{\rm{D}} \leftarrow {\rm{C}}(\Omega )} ({\bf{q}}) &= \sum\limits_l {\tilde C_{o'_1 ,o_2 ,{\bf{R}}_l ;o_1 ,o'_2 ,{\bf{R}}_n  + {\bf{R}}_l  - {\bf{R}}_m }^\Omega  ( - {\bf{R}}_m )e^{ - i{\bf{R}}_l  \cdot {\bf{q}}}},
\end{split}
\end{eqnarray}
with the Fourier transforms of the bosonic propagators,
\begin{eqnarray} \label{eqC11}
\begin{split}
\tilde P^\Omega  ({\bf{R}}_m ) &\equiv \frac{1}{{S_{BZ} }}\int {d{\bf{q}}} P^\Omega  ({\bf{q}})e^{ - i{\bf{R}}_m  \cdot {\bf{q}}} ,\\
\tilde C^\Omega  ({\bf{R}}_m ) &\equiv \frac{1}{{S_{BZ} }}\int {d{\bf{q}}} C^\Omega  ({\bf{q}})e^{ - i{\bf{R}}_m  \cdot {\bf{q}}} ,\tilde D^\Omega  ({\bf{R}}_m ) \equiv \frac{1}{{S_{BZ} }}\int {d{\bf{q}}} D^\Omega  ({\bf{q}})e^{ - i{\bf{R}}_m  \cdot {\bf{q}}} .
\end{split}
\end{eqnarray}

\end{widetext}

\bibliographystyle{apsrev}
\bibliography{SJO_PRB_2020}

\begin{thebibliography}{60}
\expandafter\ifx\csname natexlab\endcsname\relax\def\natexlab#1{#1}\fi
\expandafter\ifx\csname bibnamefont\endcsname\relax
  \def\bibnamefont#1{#1}\fi
\expandafter\ifx\csname bibfnamefont\endcsname\relax
  \def\bibfnamefont#1{#1}\fi
\expandafter\ifx\csname citenamefont\endcsname\relax
  \def\citenamefont#1{#1}\fi
\expandafter\ifx\csname url\endcsname\relax
  \def\url#1{\texttt{#1}}\fi
\expandafter\ifx\csname urlprefix\endcsname\relax\def\urlprefix{URL }\fi
\providecommand{\bibinfo}[2]{#2}
\providecommand{\eprint}[2][]{\url{#2}}

\bibitem[{\citenamefont{Cao et~al.}(2018{\natexlab{a}})\citenamefont{Cao,
  Fatemi, Demir, Fang, Tomarken, Luo, Sanchez-Yamagishi, Watanabe, Taniguchi,
  Kaxiras et~al.}}]{ref01}
\bibinfo{author}{\bibfnamefont{Y.}~\bibnamefont{Cao}},
  \bibinfo{author}{\bibfnamefont{V.}~\bibnamefont{Fatemi}},
  \bibinfo{author}{\bibfnamefont{A.}~\bibnamefont{Demir}},
  \bibinfo{author}{\bibfnamefont{S.}~\bibnamefont{Fang}},
  \bibinfo{author}{\bibfnamefont{S.~L.} \bibnamefont{Tomarken}},
  \bibinfo{author}{\bibfnamefont{J.~Y.} \bibnamefont{Luo}},
  \bibinfo{author}{\bibfnamefont{J.~D.} \bibnamefont{Sanchez-Yamagishi}},
  \bibinfo{author}{\bibfnamefont{K.}~\bibnamefont{Watanabe}},
  \bibinfo{author}{\bibfnamefont{T.}~\bibnamefont{Taniguchi}},
  \bibinfo{author}{\bibfnamefont{E.}~\bibnamefont{Kaxiras}},
  \bibinfo{author}{\bibfnamefont{R.~C.}~\bibnamefont{Ashoori}}, \bibnamefont{and}
  \bibinfo{author}{\bibfnamefont{P.}~\bibnamefont{Jarillo-Herrero}},
  \bibinfo{journal}{Nature (London)} \textbf{\bibinfo{volume}{556}}, \bibinfo{pages}{80}
  (\bibinfo{year}{2018}{\natexlab{a}}).

\bibitem[{\citenamefont{Cao et~al.}(2018{\natexlab{b}})\citenamefont{Cao,
  Fatemi, S.~Fang, Taniguchi, Kaxiras, and Jarillo-Herrero}}]{ref02}
\bibinfo{author}{\bibfnamefont{Y.}~\bibnamefont{Cao}},
  \bibinfo{author}{\bibfnamefont{V.}~\bibnamefont{Fatemi}},
  \bibinfo{author}{\bibfnamefont{K.~W.} \bibnamefont{S.~Fang}},
  \bibinfo{author}{\bibfnamefont{T.}~\bibnamefont{Taniguchi}},
  \bibinfo{author}{\bibfnamefont{E.}~\bibnamefont{Kaxiras}}, \bibnamefont{and}
  \bibinfo{author}{\bibfnamefont{P.}~\bibnamefont{Jarillo-Herrero}},
  \bibinfo{journal}{Nature (London)} \textbf{\bibinfo{volume}{556}},
  \bibinfo{pages}{43} (\bibinfo{year}{2018}{\natexlab{b}}).

\bibitem[{\citenamefont{Yankowitz et~al.}(2019)\citenamefont{Yankowitz, Chen,
  Polshyn, Watanabe, Taniguchi, Graf, Young, and Dean}}]{ref03}
\bibinfo{author}{\bibfnamefont{M.}~\bibnamefont{Yankowitz}},
  \bibinfo{author}{\bibfnamefont{S.}~\bibnamefont{Chen}},
  \bibinfo{author}{\bibfnamefont{H.}~\bibnamefont{Polshyn}},
  \bibinfo{author}{\bibfnamefont{K.}~\bibnamefont{Watanabe}},
  \bibinfo{author}{\bibfnamefont{T.}~\bibnamefont{Taniguchi}},
  \bibinfo{author}{\bibfnamefont{D.}~\bibnamefont{Graf}},
  \bibinfo{author}{\bibfnamefont{A.~F.} \bibnamefont{Young}}, \bibnamefont{and}
  \bibinfo{author}{\bibfnamefont{C.~R.} \bibnamefont{Dean}},
  \bibinfo{journal}{Science} \textbf{\bibinfo{volume}{363}},
  \bibinfo{pages}{1059} (\bibinfo{year}{2019}).

\bibitem[{\citenamefont{Xu and Balents}(2018)}]{ref04}
\bibinfo{author}{\bibfnamefont{C.~K.} \bibnamefont{Xu}} \bibnamefont{and}
  \bibinfo{author}{\bibfnamefont{L.}~\bibnamefont{Balents}},
  \bibinfo{journal}{Phys. Rev. Lett.} \textbf{\bibinfo{volume}{121}},
  \bibinfo{pages}{087001} (\bibinfo{year}{2018}).

\bibitem[{\citenamefont{Guo et~al.}(2018)\citenamefont{Guo, Zhu, Feng, and
  Scalettar}}]{ref05}
\bibinfo{author}{\bibfnamefont{H.}~\bibnamefont{Guo}},
  \bibinfo{author}{\bibfnamefont{X.}~\bibnamefont{Zhu}},
  \bibinfo{author}{\bibfnamefont{S.}~\bibnamefont{Feng}}, \bibnamefont{and}
  \bibinfo{author}{\bibfnamefont{R.~T.} \bibnamefont{Scalettar}},
  \bibinfo{journal}{Phys. Rev.} \textbf{\bibinfo{volume}{B 97}},
  \bibinfo{pages}{235453} (\bibinfo{year}{2018}).

\bibitem[{\citenamefont{Liu et~al.}(2018)\citenamefont{Liu, Zhang, Chen, and
  Yang}}]{ref06}
\bibinfo{author}{\bibfnamefont{C.~C.} \bibnamefont{Liu}},
  \bibinfo{author}{\bibfnamefont{L.~D.} \bibnamefont{Zhang}},
  \bibinfo{author}{\bibfnamefont{W.~Q.} \bibnamefont{Chen}}, \bibnamefont{and}
  \bibinfo{author}{\bibfnamefont{F.}~\bibnamefont{Yang}},
  \bibinfo{journal}{Phys. Rev. Lett.} \textbf{\bibinfo{volume}{121}},
  \bibinfo{pages}{217001} (\bibinfo{year}{2018}).

\bibitem[{\citenamefont{Kennes et~al.}(2018)\citenamefont{Kennes, Lischner, and
  Karrasch}}]{ref07}
\bibinfo{author}{\bibfnamefont{D.~M.} \bibnamefont{Kennes}},
  \bibinfo{author}{\bibfnamefont{J.}~\bibnamefont{Lischner}}, \bibnamefont{and}
  \bibinfo{author}{\bibfnamefont{C.}~\bibnamefont{Karrasch}},
  \bibinfo{journal}{Phys. Rev.} \textbf{\bibinfo{volume}{B 98}},
  \bibinfo{pages}{241407(R)} (\bibinfo{year}{2018}).

\bibitem[{\citenamefont{Lin and Nandkishore}(2018)}]{ref08}
\bibinfo{author}{\bibfnamefont{Y.-P.} \bibnamefont{Lin}} \bibnamefont{and}
  \bibinfo{author}{\bibfnamefont{R.~M.} \bibnamefont{Nandkishore}},
  \bibinfo{journal}{Phys. Rev.} \textbf{\bibinfo{volume}{B 98}},
  \bibinfo{pages}{214521} (\bibinfo{year}{2018}).

\bibitem[{\citenamefont{Huang et~al.}(2019)\citenamefont{Huang, Zhang, and
  Ma}}]{ref09}
\bibinfo{author}{\bibfnamefont{T.}~\bibnamefont{Huang}},
  \bibinfo{author}{\bibfnamefont{L.}~\bibnamefont{Zhang}}, \bibnamefont{and}
  \bibinfo{author}{\bibfnamefont{T.}~\bibnamefont{Ma}}, \bibinfo{journal}{Sci.
  Bull.} \textbf{\bibinfo{volume}{64}}, \bibinfo{pages}{310}
  (\bibinfo{year}{2019}),
  \urlprefix\url{https://doi.org/10.1016/j.scib.2019.01.026}.

\bibitem[{\citenamefont{Fidrysiak et~al.}(2018)\citenamefont{Fidrysiak,
  Zegrodnik, and Spa\l{}ek}}]{ref10}
\bibinfo{author}{\bibfnamefont{M.}~\bibnamefont{Fidrysiak}},
  \bibinfo{author}{\bibfnamefont{M.}~\bibnamefont{Zegrodnik}},
  \bibnamefont{and}
  \bibinfo{author}{\bibfnamefont{J.}~\bibnamefont{Spa\l{}ek}},
  \bibinfo{journal}{Phys. Rev.} \textbf{\bibinfo{volume}{B 98}},
  \bibinfo{pages}{085436} (\bibinfo{year}{2018}).

\bibitem[{\citenamefont{Classen et~al.}(2019)\citenamefont{Classen, Honerkamp,
  and Scherer}}]{ref57}
\bibinfo{author}{\bibfnamefont{L.}~\bibnamefont{Classen}},
  \bibinfo{author}{\bibfnamefont{C.}~\bibnamefont{Honerkamp}},
  \bibnamefont{and} \bibinfo{author}{\bibfnamefont{M.~M.}
  \bibnamefont{Scherer}}, \bibinfo{journal}{Phys. Rev.}
  \textbf{\bibinfo{volume}{B 99}}, \bibinfo{pages}{195120}
  (\bibinfo{year}{2019}).

\bibitem[{\citenamefont{Lin and Nandkishore}(2019)}]{ref58}
\bibinfo{author}{\bibfnamefont{Y.-P.} \bibnamefont{Lin}} \bibnamefont{and}
  \bibinfo{author}{\bibfnamefont{R.~M.} \bibnamefont{Nandkishore}},
  \bibinfo{journal}{Phys. Rev.} \textbf{\bibinfo{volume}{B 100}},
  \bibinfo{pages}{085136} (\bibinfo{year}{2019}).

\bibitem[{\citenamefont{Chichinadze et~al.}(2020)\citenamefont{Chichinadze,
  Classen, and Chubukov}}]{ref59}
\bibinfo{author}{\bibfnamefont{D.~V.}~\bibnamefont{Chichinadze}},
  \bibinfo{author}{\bibfnamefont{L.}~\bibnamefont{Classen}}, \bibnamefont{and}
  \bibinfo{author}{\bibfnamefont{A.~V.} \bibnamefont{Chubukov}},
  \bibinfo{journal}{Phys. Rev.} \textbf{\bibinfo{volume}{B 101}},
  \bibinfo{pages}{224513} (\bibinfo{year}{2020}).

\bibitem[{\citenamefont{Fischer et~al.}(2020)\citenamefont{Fischer, Klebl,
  Honerkamp, and Kennes}}]{ref60}
\bibinfo{author}{\bibfnamefont{A.}~\bibnamefont{Fischer}},
  \bibinfo{author}{\bibfnamefont{L.}~\bibnamefont{Klebl}},
  \bibinfo{author}{\bibfnamefont{C.}~\bibnamefont{Honerkamp}},
  \bibnamefont{and} \bibinfo{author}{\bibfnamefont{D.~M.}
  \bibnamefont{Kennes}},
  \bibinfo{journal}{Phys. Rev.} \textbf{\bibinfo{volume}{B 103}},
  \bibinfo{pages}{041103} (\bibinfo{year}{2021}).


\bibitem[{\citenamefont{Nandkishore
  et~al.}(2012{\natexlab{a}})\citenamefont{Nandkishore, Levitov, and
  Chubukov}}]{ref11}
\bibinfo{author}{\bibfnamefont{R.}~\bibnamefont{Nandkishore}},
  \bibinfo{author}{\bibfnamefont{L.~S.} \bibnamefont{Levitov}},
  \bibnamefont{and} \bibinfo{author}{\bibfnamefont{A.~V.}
  \bibnamefont{Chubukov}}, \bibinfo{journal}{Nat. Phys.}
  \textbf{\bibinfo{volume}{8}}, \bibinfo{pages}{158}
  (\bibinfo{year}{2012}{\natexlab{a}}).

\bibitem[{\citenamefont{Black-Schaffer and Honerkamp}(2014)}]{ref12}
\bibinfo{author}{\bibfnamefont{A.~M.} \bibnamefont{Black-Schaffer}}
  \bibnamefont{and}
  \bibinfo{author}{\bibfnamefont{C.}~\bibnamefont{Honerkamp}},
  \bibinfo{journal}{J. Phys.: Condens. Matter} \textbf{\bibinfo{volume}{26}},
  \bibinfo{pages}{423201} (\bibinfo{year}{2014}).

\bibitem[{\citenamefont{Black-Schaffer and Doniach}(2007)}]{ref13}
\bibinfo{author}{\bibfnamefont{A.~M.} \bibnamefont{Black-Schaffer}}
  \bibnamefont{and} \bibinfo{author}{\bibfnamefont{S.}~\bibnamefont{Doniach}},
  \bibinfo{journal}{Phys. Rev.} \textbf{\bibinfo{volume}{B 75}},
  \bibinfo{pages}{134512} (\bibinfo{year}{2007}).

\bibitem[{\citenamefont{Honerkamp}(2008)}]{ref14}
\bibinfo{author}{\bibfnamefont{C.}~\bibnamefont{Honerkamp}},
  \bibinfo{journal}{Phys. Rev. Lett.} \textbf{\bibinfo{volume}{100}},
  \bibinfo{pages}{146404} (\bibinfo{year}{2008}).

\bibitem[{\citenamefont{Pathak et~al.}(2010)\citenamefont{Pathak, Shenoy, and
  Baskaran}}]{ref15}
\bibinfo{author}{\bibfnamefont{S.}~\bibnamefont{Pathak}},
  \bibinfo{author}{\bibfnamefont{V.~B.} \bibnamefont{Shenoy}},
  \bibnamefont{and} \bibinfo{author}{\bibfnamefont{G.}~\bibnamefont{Baskaran}},
  \bibinfo{journal}{Phys. Rev.} \textbf{\bibinfo{volume}{B 81}},
  \bibinfo{pages}{085431} (\bibinfo{year}{2010}).

\bibitem[{\citenamefont{Ma et~al.}(2011)\citenamefont{Ma, Huang, Hu, and
  Lin}}]{ref16}
\bibinfo{author}{\bibfnamefont{T.}~\bibnamefont{Ma}},
  \bibinfo{author}{\bibfnamefont{Z.}~\bibnamefont{Huang}},
  \bibinfo{author}{\bibfnamefont{F.}~\bibnamefont{Hu}}, \bibnamefont{and}
  \bibinfo{author}{\bibfnamefont{H.-Q.} \bibnamefont{Lin}},
  \bibinfo{journal}{Phys. Rev.} \textbf{\bibinfo{volume}{B 84}},
  \bibinfo{pages}{121410(R)} (\bibinfo{year}{2011}).

\bibitem[{\citenamefont{Wu et~al.}(2013)\citenamefont{Wu, Scherer, Honerkamp,
  and {Le Hur}}}]{ref17}
\bibinfo{author}{\bibfnamefont{W.}~\bibnamefont{Wu}},
  \bibinfo{author}{\bibfnamefont{M.~M.} \bibnamefont{Scherer}},
  \bibinfo{author}{\bibfnamefont{C.}~\bibnamefont{Honerkamp}},
  \bibnamefont{and} \bibinfo{author}{\bibfnamefont{K.}~\bibnamefont{{Le Hur}}},
  \bibinfo{journal}{Phys. Rev.} \textbf{\bibinfo{volume}{B 87}},
  \bibinfo{pages}{094521} (\bibinfo{year}{2013}).

\bibitem[{\citenamefont{Schulz}(1987)}]{ref18}
\bibinfo{author}{\bibfnamefont{H.~J.} \bibnamefont{Schulz}},
  \bibinfo{journal}{Europhys. Lett.} \textbf{\bibinfo{volume}{4}},
  \bibinfo{pages}{609} (\bibinfo{year}{1987}).

\bibitem[{\citenamefont{Dzyaloshinskii}(1987)}]{ref19}
\bibinfo{author}{\bibfnamefont{I.~E.} \bibnamefont{Dzyaloshinskii}},
  \bibinfo{journal}{Sov. Phys. JETP} \textbf{\bibinfo{volume}{66}},
  \bibinfo{pages}{848} (\bibinfo{year}{1987}).

\bibitem[{\citenamefont{Furukawa et~al.}(1998)\citenamefont{Furukawa, Rice, and
  Salmhofer}}]{ref20}
\bibinfo{author}{\bibfnamefont{N.}~\bibnamefont{Furukawa}},
  \bibinfo{author}{\bibfnamefont{T.~M.} \bibnamefont{Rice}}, \bibnamefont{and}
  \bibinfo{author}{\bibfnamefont{M.}~\bibnamefont{Salmhofer}},
  \bibinfo{journal}{Phys. Rev. Lett.} \textbf{\bibinfo{volume}{81}},
  \bibinfo{pages}{3195} (\bibinfo{year}{1998}).

\bibitem[{\citenamefont{Kotov et~al.}(2012)\citenamefont{Kotov, Uchoa, Pereira,
  Guinea, and {Castro Neto}}}]{ref21}
\bibinfo{author}{\bibfnamefont{V.~N.} \bibnamefont{Kotov}},
  \bibinfo{author}{\bibfnamefont{B.}~\bibnamefont{Uchoa}},
  \bibinfo{author}{\bibfnamefont{V.~M.} \bibnamefont{Pereira}},
  \bibinfo{author}{\bibfnamefont{F.}~\bibnamefont{Guinea}}, \bibnamefont{and}
  \bibinfo{author}{\bibfnamefont{A.~H.} \bibnamefont{{Castro Neto}}},
  \bibinfo{journal}{Rev. Mod. Phys.} \textbf{\bibinfo{volume}{84}},
  \bibinfo{pages}{1067} (\bibinfo{year}{2012}).

\bibitem[{\citenamefont{Valenzuela and Vozmediano}(2008)}]{ref29}
\bibinfo{author}{\bibfnamefont{B.}~\bibnamefont{Valenzuela}} \bibnamefont{and}
  \bibinfo{author}{\bibfnamefont{M.~A.~H.} \bibnamefont{Vozmediano}},
  \bibinfo{journal}{New J. Phys.} \textbf{\bibinfo{volume}{10}},
  \bibinfo{pages}{113009} (\bibinfo{year}{2008}).

\bibitem[{\citenamefont{Li}(2012)}]{ref31}
\bibinfo{author}{\bibfnamefont{T.}~\bibnamefont{Li}},
  \bibinfo{journal}{Europhys. Lett.} \textbf{\bibinfo{volume}{97}},
  \bibinfo{pages}{37001} (\bibinfo{year}{2012}).

\bibitem[{\citenamefont{Nandkishore
  et~al.}(2012{\natexlab{b}})\citenamefont{Nandkishore, Chern, and
  Chubukov}}]{ref34}
\bibinfo{author}{\bibfnamefont{R.}~\bibnamefont{Nandkishore}},
  \bibinfo{author}{\bibfnamefont{G.~W.} \bibnamefont{Chern}}, \bibnamefont{and}
  \bibinfo{author}{\bibfnamefont{A.}~\bibnamefont{Chubukov}},
  \bibinfo{journal}{Phys. Rev. Lett.} \textbf{\bibinfo{volume}{108}},
  \bibinfo{pages}{227204} (\bibinfo{year}{2012}{\natexlab{b}}).

\bibitem[{\citenamefont{Gonz\'{a}lez}(2008)}]{ref22}
\bibinfo{author}{\bibfnamefont{J.}~\bibnamefont{Gonz\'{a}lez}},
  \bibinfo{journal}{Phys. Rev.} \textbf{\bibinfo{volume}{B 78}},
  \bibinfo{pages}{205431} (\bibinfo{year}{2008}).

\bibitem[{\citenamefont{Makogon et~al.}(2011)\citenamefont{Makogon, van
  Gelderen, Rold\'{a}n, and Smith}}]{ref33}
\bibinfo{author}{\bibfnamefont{D.}~\bibnamefont{Makogon}},
  \bibinfo{author}{\bibfnamefont{R.}~\bibnamefont{van Gelderen}},
  \bibinfo{author}{\bibfnamefont{R.}~\bibnamefont{Rold\'{a}n}},
  \bibnamefont{and} \bibinfo{author}{\bibfnamefont{C.~M.} \bibnamefont{Smith}},
  \bibinfo{journal}{Phys. Rev.} \textbf{\bibinfo{volume}{B 84}},
  \bibinfo{pages}{125404} (\bibinfo{year}{2011}).

\bibitem[{\citenamefont{Ying and Wessel}(2018)}]{ref25}
\bibinfo{author}{\bibfnamefont{T.}~\bibnamefont{Ying}} \bibnamefont{and}
  \bibinfo{author}{\bibfnamefont{S.}~\bibnamefont{Wessel}},
  \bibinfo{journal}{Phys. Rev.} \textbf{\bibinfo{volume}{B 97}},
  \bibinfo{pages}{075127} (\bibinfo{year}{2018}).

\bibitem[{\citenamefont{Jiang et~al.}(2014)\citenamefont{Jiang, Mesaros, and
  Ran}}]{ref32}
\bibinfo{author}{\bibfnamefont{S.}~\bibnamefont{Jiang}},
  \bibinfo{author}{\bibfnamefont{A.}~\bibnamefont{Mesaros}}, \bibnamefont{and}
  \bibinfo{author}{\bibfnamefont{Y.}~\bibnamefont{Ran}},
  \bibinfo{journal}{Phys. Rev.} \textbf{\bibinfo{volume}{X 4}},
  \bibinfo{pages}{031040} (\bibinfo{year}{2014}).

\bibitem[{\citenamefont{Gu et~al.}(2013)\citenamefont{Gu, Jiang, Sheng, Yao,
  Balents, and Wen}}]{ref24}
\bibinfo{author}{\bibfnamefont{Z.-C.} \bibnamefont{Gu}},
  \bibinfo{author}{\bibfnamefont{H.-C.} \bibnamefont{Jiang}},
  \bibinfo{author}{\bibfnamefont{D.~N.} \bibnamefont{Sheng}},
  \bibinfo{author}{\bibfnamefont{H.}~\bibnamefont{Yao}},
  \bibinfo{author}{\bibfnamefont{L.}~\bibnamefont{Balents}}, \bibnamefont{and}
  \bibinfo{author}{\bibfnamefont{X.-G.} \bibnamefont{Wen}},
  \bibinfo{journal}{Phys. Rev.} \textbf{\bibinfo{volume}{B 88}},
  \bibinfo{pages}{155112} (\bibinfo{year}{2013}).

\bibitem[{\citenamefont{Faye et~al.}(2015)\citenamefont{Faye, Sahebsara, and
  S\'{e}n\'{e}chal}}]{ref27}
\bibinfo{author}{\bibfnamefont{J.~P.~L.} \bibnamefont{Faye}},
  \bibinfo{author}{\bibfnamefont{P.}~\bibnamefont{Sahebsara}},
  \bibnamefont{and}
  \bibinfo{author}{\bibfnamefont{D.}~\bibnamefont{S\'{e}n\'{e}chal}},
  \bibinfo{journal}{Phys. Rev.} \textbf{\bibinfo{volume}{B 92}},
  \bibinfo{pages}{085121} (\bibinfo{year}{2015}).

\bibitem[{\citenamefont{Kiesel et~al.}(2012)\citenamefont{Kiesel, Platt, Hanke,
  Abanin, and Thomale}}]{ref23}
\bibinfo{author}{\bibfnamefont{M.~L.} \bibnamefont{Kiesel}},
  \bibinfo{author}{\bibfnamefont{C.}~\bibnamefont{Platt}},
  \bibinfo{author}{\bibfnamefont{W.}~\bibnamefont{Hanke}},
  \bibinfo{author}{\bibfnamefont{D.~A.} \bibnamefont{Abanin}},
  \bibnamefont{and} \bibinfo{author}{\bibfnamefont{R.}~\bibnamefont{Thomale}},
  \bibinfo{journal}{Phys. Rev.} \textbf{\bibinfo{volume}{B 86}},
  \bibinfo{pages}{020507(R)} (\bibinfo{year}{2012}).

\bibitem[{\citenamefont{Wang et~al.}(2012)\citenamefont{Wang, Xiang, Wang,
  Wang, Yang, and Lee}}]{ref26}
\bibinfo{author}{\bibfnamefont{W.-S.} \bibnamefont{Wang}},
  \bibinfo{author}{\bibfnamefont{Y.-Y.} \bibnamefont{Xiang}},
  \bibinfo{author}{\bibfnamefont{Q.-H.} \bibnamefont{Wang}},
  \bibinfo{author}{\bibfnamefont{F.}~\bibnamefont{Wang}},
  \bibinfo{author}{\bibfnamefont{F.}~\bibnamefont{Yang}}, \bibnamefont{and}
  \bibinfo{author}{\bibfnamefont{D.-H.} \bibnamefont{Lee}},
  \bibinfo{journal}{Phys. Rev.} \textbf{\bibinfo{volume}{B 85}},
  \bibinfo{pages}{035414} (\bibinfo{year}{2012}).

\bibitem[{\citenamefont{Xu et~al.}(2016)\citenamefont{Xu, Wessel, and
  Meng}}]{ref28}
\bibinfo{author}{\bibfnamefont{X.~Y.} \bibnamefont{Xu}},
  \bibinfo{author}{\bibfnamefont{S.}~\bibnamefont{Wessel}}, \bibnamefont{and}
  \bibinfo{author}{\bibfnamefont{Z.~Y.} \bibnamefont{Meng}},
  \bibinfo{journal}{Phys. Rev.} \textbf{\bibinfo{volume}{B 94}},
  \bibinfo{pages}{115105} (\bibinfo{year}{2016}).

\bibitem[{\citenamefont{Lamas et~al.}(2009)\citenamefont{Lamas, Cabra, and
  Grandi}}]{ref30}
\bibinfo{author}{\bibfnamefont{C.~A.} \bibnamefont{Lamas}},
  \bibinfo{author}{\bibfnamefont{D.~C.} \bibnamefont{Cabra}}, \bibnamefont{and}
  \bibinfo{author}{\bibfnamefont{N.}~\bibnamefont{Grandi}},
  \bibinfo{journal}{Phys. Rev.} \textbf{\bibinfo{volume}{B 80}},
  \bibinfo{pages}{075108} (\bibinfo{year}{2009}).

\bibitem[{\citenamefont{O et~al.}(2019)\citenamefont{O, Kim, Rim, Pak, and
  Im}}]{ref35}
\bibinfo{author}{\bibfnamefont{S.-J.} \bibnamefont{O}},
  \bibinfo{author}{\bibfnamefont{Y.-H.} \bibnamefont{Kim}},
  \bibinfo{author}{\bibfnamefont{H.-Y.} \bibnamefont{Rim}},
  \bibinfo{author}{\bibfnamefont{H.-C.} \bibnamefont{Pak}}, \bibnamefont{and}
  \bibinfo{author}{\bibfnamefont{S.-J.} \bibnamefont{Im}},
  \bibinfo{journal}{Phys. Rev.} \textbf{\bibinfo{volume}{B 99}},
  \bibinfo{pages}{245140} (\bibinfo{year}{2019}).

\bibitem[{\citenamefont{Lichtenstein et~al.}(2017)\citenamefont{Lichtenstein,
  de~la Pe\~{n}a, Rohe, Napoli, Honerkamp, and Maier}}]{ref36}
\bibinfo{author}{\bibfnamefont{J.}~\bibnamefont{Lichtenstein}},
  \bibinfo{author}{\bibfnamefont{D.~S.} \bibnamefont{de~la Pe\~{n}a}},
  \bibinfo{author}{\bibfnamefont{D.}~\bibnamefont{Rohe}},
  \bibinfo{author}{\bibfnamefont{E.~D.} \bibnamefont{Napoli}},
  \bibinfo{author}{\bibfnamefont{C.}~\bibnamefont{Honerkamp}},
  \bibnamefont{and} \bibinfo{author}{\bibfnamefont{S.~A.} \bibnamefont{Maier}},
  \bibinfo{journal}{Comput. Phys. Commun.} \textbf{\bibinfo{volume}{213}},
  \bibinfo{pages}{100} (\bibinfo{year}{2017}).

\bibitem[{\citenamefont{{Castro Neto} et~al.}(2009)\citenamefont{{Castro Neto},
  Guinea, Peres, Novoselov, and Geim}}]{ref37}
\bibinfo{author}{\bibfnamefont{A.~H.} \bibnamefont{{Castro Neto}}},
  \bibinfo{author}{\bibfnamefont{F.}~\bibnamefont{Guinea}},
  \bibinfo{author}{\bibfnamefont{N.~M.~R.} \bibnamefont{Peres}},
  \bibinfo{author}{\bibfnamefont{K.~S.} \bibnamefont{Novoselov}},
  \bibnamefont{and} \bibinfo{author}{\bibfnamefont{A.~K.} \bibnamefont{Geim}},
  \bibinfo{journal}{Rev. Mod. Phys.} \textbf{\bibinfo{volume}{81}},
  \bibinfo{pages}{109} (\bibinfo{year}{2009}).

\bibitem[{\citenamefont{Kopietz et~al.}(2010)\citenamefont{Kopietz, Bartosch,
  and Sch\"{u}tz}}]{ref38}
\bibinfo{author}{\bibfnamefont{P.}~\bibnamefont{Kopietz}},
  \bibinfo{author}{\bibfnamefont{L.}~\bibnamefont{Bartosch}}, \bibnamefont{and}
  \bibinfo{author}{\bibfnamefont{F.}~\bibnamefont{Sch\"{u}tz}},
  \emph{\bibinfo{title}{Introduction to the Functional Renormalization Group}}
  (\bibinfo{publisher}{Springer, Berlin}, \bibinfo{year}{2010}).

\bibitem[{\citenamefont{Metzner et~al.}(2012)\citenamefont{Metzner, Salmhofer,
  Honerkamp, Meden, and Sch\"{o}nhammer}}]{ref39}
\bibinfo{author}{\bibfnamefont{W.}~\bibnamefont{Metzner}},
  \bibinfo{author}{\bibfnamefont{M.}~\bibnamefont{Salmhofer}},
  \bibinfo{author}{\bibfnamefont{C.}~\bibnamefont{Honerkamp}},
  \bibinfo{author}{\bibfnamefont{V.}~\bibnamefont{Meden}}, \bibnamefont{and}
  \bibinfo{author}{\bibfnamefont{K.}~\bibnamefont{Sch\"{o}nhammer}},
  \bibinfo{journal}{Rev. Mod. Phys.} \textbf{\bibinfo{volume}{84}},
  \bibinfo{pages}{299} (\bibinfo{year}{2012}).

\bibitem[{\citenamefont{Platt et~al.}(2013)\citenamefont{Platt, Hanke, and
  Thomale}}]{ref40}
\bibinfo{author}{\bibfnamefont{C.}~\bibnamefont{Platt}},
  \bibinfo{author}{\bibfnamefont{W.}~\bibnamefont{Hanke}}, \bibnamefont{and}
  \bibinfo{author}{\bibfnamefont{R.}~\bibnamefont{Thomale}},
  \bibinfo{journal}{Adv. Phys.} \textbf{\bibinfo{volume}{62}},
  \bibinfo{pages}{453} (\bibinfo{year}{2013}).

\bibitem[{\citenamefont{Classen et~al.}(2014)\citenamefont{Classen, Scherer,
  and Honerkamp}}]{ref54}
\bibinfo{author}{\bibfnamefont{L.}~\bibnamefont{Classen}},
  \bibinfo{author}{\bibfnamefont{M.~M.} \bibnamefont{Scherer}},
  \bibnamefont{and}
  \bibinfo{author}{\bibfnamefont{C.}~\bibnamefont{Honerkamp}},
  \bibinfo{journal}{Phys. Rev.} \textbf{\bibinfo{volume}{B 90}},
  \bibinfo{pages}{035122} (\bibinfo{year}{2014}).

\bibitem[{\citenamefont{Husemann and Salmhofer}(2009)}]{ref41}
\bibinfo{author}{\bibfnamefont{C.}~\bibnamefont{Husemann}} \bibnamefont{and}
  \bibinfo{author}{\bibfnamefont{M.}~\bibnamefont{Salmhofer}},
  \bibinfo{journal}{Phys. Rev.} \textbf{\bibinfo{volume}{B 79}},
  \bibinfo{pages}{195125} (\bibinfo{year}{2009}).

\bibitem[{\citenamefont{Schober et~al.}(2018)\citenamefont{Schober, Ehrlich,
  Reckling, and Honerkamp}}]{ref42}
\bibinfo{author}{\bibfnamefont{G.~A.~H.} \bibnamefont{Schober}},
  \bibinfo{author}{\bibfnamefont{J.}~\bibnamefont{Ehrlich}},
  \bibinfo{author}{\bibfnamefont{T.}~\bibnamefont{Reckling}}, \bibnamefont{and}
  \bibinfo{author}{\bibfnamefont{C.}~\bibnamefont{Honerkamp}},
  \bibinfo{journal}{Frontiers Phys.} \textbf{\bibinfo{volume}{6}},
  \bibinfo{pages}{32} (\bibinfo{year}{2018}),
  \urlprefix\url{https://doi.org/10.3389/fphy.2018.00032}.

\bibitem[{\citenamefont{de~la Pe\~{n}a
  et~al.}(2017{\natexlab{a}})\citenamefont{de~la Pe\~{n}a, Lichtenstein, and
  Honerkamp}}]{ref43}
\bibinfo{author}{\bibfnamefont{D.~S.} \bibnamefont{de~la Pe\~{n}a}},
  \bibinfo{author}{\bibfnamefont{J.}~\bibnamefont{Lichtenstein}},
  \bibnamefont{and}
  \bibinfo{author}{\bibfnamefont{C.}~\bibnamefont{Honerkamp}},
  \bibinfo{journal}{Phys. Rev.} \textbf{\bibinfo{volume}{B 95}},
  \bibinfo{pages}{085143} (\bibinfo{year}{2017}{\natexlab{a}}).

\bibitem[{\citenamefont{de~la Pe\~{n}a
  et~al.}(2017{\natexlab{b}})\citenamefont{de~la Pe\~{n}a, Lichtenstein,
  Honerkamp, and Scherer}}]{ref44}
\bibinfo{author}{\bibfnamefont{D.~S.} \bibnamefont{de~la Pe\~{n}a}},
  \bibinfo{author}{\bibfnamefont{J.}~\bibnamefont{Lichtenstein}},
  \bibinfo{author}{\bibfnamefont{C.}~\bibnamefont{Honerkamp}},
  \bibnamefont{and} \bibinfo{author}{\bibfnamefont{M.~M.}
  \bibnamefont{Scherer}}, \bibinfo{journal}{Phys. Rev.}
  \textbf{\bibinfo{volume}{B 96}}, \bibinfo{pages}{205155}
  (\bibinfo{year}{2017}{\natexlab{b}}).

\bibitem[{\citenamefont{Salmhofer and Honerkamp}(2001)}]{ref45}
\bibinfo{author}{\bibfnamefont{M.}~\bibnamefont{Salmhofer}} \bibnamefont{and}
  \bibinfo{author}{\bibfnamefont{C.}~\bibnamefont{Honerkamp}},
  \bibinfo{journal}{Prog. Theor. Phys.} \textbf{\bibinfo{volume}{105}},
  \bibinfo{pages}{1} (\bibinfo{year}{2001}).

\bibitem[{\citenamefont{Honerkamp et~al.}(2001)\citenamefont{Honerkamp,
  Salmhofer, Furukawa, and Rice}}]{ref46}
\bibinfo{author}{\bibfnamefont{C.}~\bibnamefont{Honerkamp}},
  \bibinfo{author}{\bibfnamefont{M.}~\bibnamefont{Salmhofer}},
  \bibinfo{author}{\bibfnamefont{N.}~\bibnamefont{Furukawa}}, \bibnamefont{and}
  \bibinfo{author}{\bibfnamefont{T.~M.} \bibnamefont{Rice}},
  \bibinfo{journal}{Phys. Rev.} \textbf{\bibinfo{volume}{B 63}},
  \bibinfo{pages}{035109} (\bibinfo{year}{2001}).

\bibitem[{Note1()}]{Note1}
\bibinfo{note}{There is an error in expression for $J^{{\protect \rm
  {ph,cr}}}$ in Eq. (5) of Ref. \onlinecite {ref35}, which was revised in Eq. (\ref
  {eq06}) of this paper. In addition, there is an another error in expression
  for $T_{ob}({\protect \bf {k}})$ in Eq. (43) of the reference, which should
  be corrected as $T_{ob}({\protect \bf {k}}) = \left ({\protect \sqrt 2}\right
  )^{-1} \left (\begin {array}{cc} \protect \frac {d({\protect \bf
  {k}})}{|d({\protect \bf {k}})|} \hskip 1em\relax \protect \frac {d({\protect
  \bf {k}})}{|d({\protect \bf {k}})|}\\ -1 {\kern 23pt} 1{\kern 6pt} \end
  {array}\right )$.}

\bibitem[{\citenamefont{Wang et~al.}(2014)\citenamefont{Wang, Eberlein, and
  Metzner}}]{ref48}
\bibinfo{author}{\bibfnamefont{J.}~\bibnamefont{Wang}},
  \bibinfo{author}{\bibfnamefont{A.}~\bibnamefont{Eberlein}}, \bibnamefont{and}
  \bibinfo{author}{\bibfnamefont{W.}~\bibnamefont{Metzner}},
  \bibinfo{journal}{Phys. Rev.} \textbf{\bibinfo{volume}{B 89}},
  \bibinfo{pages}{121116(R)} (\bibinfo{year}{2014}).

\bibitem[{Note2()}]{Note2}
\bibinfo{note}{The coexistence phases in Fig. \ref {fig3} have not been
  rigorously verified. They were identified by comparing the strengths of
  divergences for dominant eigenmodes of the $W$ matrices in Eq. (\ref {eq23}).
  Concretely, the notation \protect \emph {Coexistence of strong A and weak B}
  means that, at the critical scale, the eigenvalue of $W$ matrix, associated
  with the A phase, $\lambda _{\protect \rm {A}}$ is $2 \sim 5$ times larger
  than that associated with the B phase, $\lambda _{\protect \rm {B}}$ (i.e.,
  $\lambda _{\protect \rm {A}}/5 < \lambda _{\protect \rm {B}} \le \lambda
  _{\protect \rm {A}}/2$). The notation \protect \emph {Coexistence of A and B
  with similar strengths} means the relation $\lambda _{\protect \rm {A}}/2 <
  \lambda _{\protect \rm {B}} \le \lambda _{\protect \rm {A}}$ at the critical
  scale, while \protect \emph {Coexistence of A and B, with similar strengths,
  and weak C} means $\lambda _{\protect \rm {A}}/2 < \lambda _{\protect \rm
  {B}} \le \lambda _{\protect \rm {A}}$ as well as $\lambda _{\protect \rm
  {A}}/5 < \lambda _{\protect \rm {C}} \le \lambda _{\protect \rm {A}}/2$.}

\bibitem[{\citenamefont{Kohn and Luttinger}(1965)}]{ref55}
\bibinfo{author}{\bibfnamefont{W.}~\bibnamefont{Kohn}} \bibnamefont{and}
  \bibinfo{author}{\bibfnamefont{J.~M.} \bibnamefont{Luttinger}},
  \bibinfo{journal}{Phys. Rev. Lett.} \textbf{\bibinfo{volume}{15}},
  \bibinfo{pages}{524} (\bibinfo{year}{1965}).

\bibitem[{\citenamefont{Ludbrook et~al.}(2015)\citenamefont{Ludbrook, Levy,
  Nigge, Zonno, Schneider, Dvorak, Veenstra, Zhdanovich, Wong, Dosanjh
  et~al.}}]{ref49}
\bibinfo{author}{\bibfnamefont{B.~M.} \bibnamefont{Ludbrook}},
  \bibinfo{author}{\bibfnamefont{G.}~\bibnamefont{Levy}},
  \bibinfo{author}{\bibfnamefont{P.}~\bibnamefont{Nigge}},
  \bibinfo{author}{\bibfnamefont{M.}~\bibnamefont{Zonno}},
  \bibinfo{author}{\bibfnamefont{M.}~\bibnamefont{Schneider}},
  \bibinfo{author}{\bibfnamefont{D.~J.} \bibnamefont{Dvorak}},
  \bibinfo{author}{\bibfnamefont{C.~N.} \bibnamefont{Veenstra}},
  \bibinfo{author}{\bibfnamefont{S.}~\bibnamefont{Zhdanovich}},
  \bibinfo{author}{\bibfnamefont{D.}~\bibnamefont{Wong}},
  \bibinfo{author}{\bibfnamefont{P.}~\bibnamefont{Dosanjh}},
  \bibinfo{author}{\bibfnamefont{C.}~\bibnamefont{Stra{\ss}er}},
  \bibinfo{author}{\bibfnamefont{A.}~\bibnamefont{St\"{o}hr}},
  \bibinfo{author}{\bibfnamefont{S.}~\bibnamefont{Forti}},
  \bibinfo{author}{\bibfnamefont{C.~R.}~\bibnamefont{Ast}},
  \bibinfo{author}{\bibfnamefont{U.}~\bibnamefont{Starke}}, \bibnamefont{and}
  \bibinfo{author}{\bibfnamefont{A.}~\bibnamefont{Damascelli}},  
  \bibinfo{journal}{Proc. Natl. Acad. Sci. USA} \textbf{\bibinfo{volume}{112}},
  \bibinfo{pages}{11795} (\bibinfo{year}{2015}).

\bibitem[{\citenamefont{Chapman et~al.}(2016)\citenamefont{Chapman, Su, Howard,
  Kundys, Grigorenko, Guinea, Geim, Grigorieva, and Nair}}]{ref50}
\bibinfo{author}{\bibfnamefont{J.}~\bibnamefont{Chapman}},
  \bibinfo{author}{\bibfnamefont{Y.}~\bibnamefont{Su}},
  \bibinfo{author}{\bibfnamefont{C.~A.} \bibnamefont{Howard}},
  \bibinfo{author}{\bibfnamefont{D.}~\bibnamefont{Kundys}},
  \bibinfo{author}{\bibfnamefont{A.}~\bibnamefont{Grigorenko}},
  \bibinfo{author}{\bibfnamefont{F.}~\bibnamefont{Guinea}},
  \bibinfo{author}{\bibfnamefont{A.~K.} \bibnamefont{Geim}},
  \bibinfo{author}{\bibfnamefont{I.~V.} \bibnamefont{Grigorieva}},
  \bibnamefont{and} \bibinfo{author}{\bibfnamefont{R.~R.} \bibnamefont{Nair}},
  \bibinfo{journal}{Sci. Rep.} \textbf{\bibinfo{volume}{6}},
  \bibinfo{pages}{23254} (\bibinfo{year}{2016}).

\bibitem[{\citenamefont{Tonnoir et~al.}(2013)\citenamefont{Tonnoir, Kimouche,
  Coraux, Magaud, Delsol, Gilles, and Chapelier}}]{ref51}
\bibinfo{author}{\bibfnamefont{C.}~\bibnamefont{Tonnoir}},
  \bibinfo{author}{\bibfnamefont{A.}~\bibnamefont{Kimouche}},
  \bibinfo{author}{\bibfnamefont{J.}~\bibnamefont{Coraux}},
  \bibinfo{author}{\bibfnamefont{L.}~\bibnamefont{Magaud}},
  \bibinfo{author}{\bibfnamefont{B.}~\bibnamefont{Delsol}},
  \bibinfo{author}{\bibfnamefont{B.}~\bibnamefont{Gilles}}, \bibnamefont{and}
  \bibinfo{author}{\bibfnamefont{C.}~\bibnamefont{Chapelier}},
  \bibinfo{journal}{Phys. Rev. Lett.} \textbf{\bibinfo{volume}{111}},
  \bibinfo{pages}{246805} (\bibinfo{year}{2013}).

\bibitem[{\citenamefont{Bernardo et~al.}(2017)\citenamefont{Bernardo, Millo,
  Barbone, Alpern, Kalcheim, Sassi, Ott, Fazio, Yoon, Amado et~al.}}]{ref52}
\bibinfo{author}{\bibfnamefont{A.~D.} \bibnamefont{Bernardo}},
  \bibinfo{author}{\bibfnamefont{O.}~\bibnamefont{Millo}},
  \bibinfo{author}{\bibfnamefont{M.}~\bibnamefont{Barbone}},
  \bibinfo{author}{\bibfnamefont{H.}~\bibnamefont{Alpern}},
  \bibinfo{author}{\bibfnamefont{Y.}~\bibnamefont{Kalcheim}},
  \bibinfo{author}{\bibfnamefont{U.}~\bibnamefont{Sassi}},
  \bibinfo{author}{\bibfnamefont{A.~K.} \bibnamefont{Ott}},
  \bibinfo{author}{\bibfnamefont{D.~D.} \bibnamefont{Fazio}},
  \bibinfo{author}{\bibfnamefont{D.}~\bibnamefont{Yoon}},
  \bibinfo{author}{\bibfnamefont{M.}~\bibnamefont{Amado}},
  \bibinfo{author}{\bibfnamefont{A.~C.} \bibnamefont{Ferrari}},
  \bibinfo{author}{\bibfnamefont{J.} \bibnamefont{Linder}}, \bibnamefont{and}
  \bibinfo{author}{\bibfnamefont{J.~W.~A.} \bibnamefont{Robinson}},
  \bibinfo{journal}{Nat. Commun.} \textbf{\bibinfo{volume}{8}},
  \bibinfo{pages}{14024} (\bibinfo{year}{2017}).

\bibitem[{\citenamefont{Wehling et~al.}(2011)\citenamefont{Wehling,
  \c{S}a\c{s}\i{}o\u{g}lu, Friedrich, Lichtenstein, Katsnelson, and
  Bl\"{u}gel}}]{ref53}
\bibinfo{author}{\bibfnamefont{T.~O.} \bibnamefont{Wehling}},
  \bibinfo{author}{\bibfnamefont{E.}~\bibnamefont{\c{S}a\c{s}\i{}o\u{g}lu}},
  \bibinfo{author}{\bibfnamefont{C.}~\bibnamefont{Friedrich}},
  \bibinfo{author}{\bibfnamefont{A.~I.} \bibnamefont{Lichtenstein}},
  \bibinfo{author}{\bibfnamefont{M.~I.} \bibnamefont{Katsnelson}},
  \bibnamefont{and}
  \bibinfo{author}{\bibfnamefont{S.}~\bibnamefont{Bl\"{u}gel}},
  \bibinfo{journal}{Phys. Rev. Lett.} \textbf{\bibinfo{volume}{106}},
  \bibinfo{pages}{236805} (\bibinfo{year}{2011}).

\end{thebibliography}

\end{document}